\documentclass{aa}

\makeatletter

\renewcommand*\aa@pageof{, page \thepage{} of \pageref*{LastPage}}

\usepackage{natbib}
\bibpunct{(}{)}{;}{a}{}{,}%
\def\bibfont{\aa@bibliographyfont}%
\makeatother

\usepackage{graphicx}
\usepackage{newtxtext}
\usepackage[smallerops,cmbraces]{newtxmath}
\usepackage{xcolor}
\definecolor{xlinkcolor}{cmyk}{1,1,0,0}
\usepackage[
  bookmarks=true,         % show bookmarks bar?/ Changed March 22, 2019 for improved accessibility
  pdfnewwindow=true,      % links in new window
  colorlinks=true,        % false: boxed links; true: colored links
  linkcolor=xlinkcolor,   % color of internal links
  citecolor=xlinkcolor,   % color of links to bibliography
  filecolor=xlinkcolor,   % color of file links
  urlcolor=xlinkcolor,    % color of external links
  final=true,
]{hyperref}

\usepackage{bm} % bold math
\usepackage{textgreek} % upright greek letters

\usepackage[capitalize]{cleveref} % easy referencing with \cref{fig:name} -> Fig.~X
\crefname{section}{Sect.}{Sects.}
\creflabelformat{equation}{#2#1#3} % equations without parentheses
\crefname{enumi}{item}{items} % items in lower case
\newcommand\appref[1]{Appendix~\ref{#1}}

\usepackage{siunitx} % proper unit typography
\DeclareSIUnit[number-unit-product = ]\percent{\char`\%} % no space before percent

\usepackage[normalem]{ulem} % sout
\usepackage{csquotes} % provides \enquote for quotation marks
\definecolor{blackberry}{HTML}{8D1D75}

\newcommand*{\code}[1]{\texttt{#1}} % wrapper for monospace font
\newcommand*{\ngc}[1]{NGC\,#1}
\newcommand*{\m}[1]{M\,#1}

\newcommand*{\OIII}{[O\,\textsc{iii}]} % [OIII]
\newcommand*{\HII}{H\,\textsc{ii}} % [HII]
\newcommand*{\Hbeta}{\ensuremath{\mathrm{H}\beta}} % Hβ

\DeclareSIUnit\parsec{pc}
\DeclareSIUnit\dex{dex}
\DeclareSIUnit\mag{mag}
\DeclareSIUnit\h{\mathnormal{h}}
\DeclareSIUnit\year{yr}
\DeclareSIUnit\years{yrs}
\DeclareSIUnit\arcsec{arcsec}
\DeclareSIUnit\arcmin{arcmin}
\DeclareSIUnit\Msun{M_\odot}
\DeclareSIUnit\Rsun{R_\odot}
\DeclareSIUnit\Lsun{L_\odot}
\DeclareSIUnit\Rvir{\mathnormal{R}_\mathrm{vir}}
\DeclareSIUnit\Rhalf{\mathnormal{R}_{1/2}}
\DeclareSIUnit\erg{erg}
\DeclareSIUnit\angstrom{\text{Å}}

\newcommand*{\Msun}{\ensuremath{\mathrm{M}_\odot}} % solar mass
\newcommand*{\Rsun}{\ensuremath{\mathrm{R}_\odot}} % solar radius
\newcommand*{\Lsun}{\ensuremath{\mathrm{L}_\odot}} % solar luminosity
\newcommand*{\Rvir}{\ensuremath{R_\mathrm{vir}}} % virial radius
\newcommand*{\Rhalf}{\ensuremath{R_{1/2}}} % half-mass radius

\begin{document}

\title{The PICS Project}
\subtitle{I. The impact of metallicity and helium abundance on the bright end of the planetary nebula luminosity function}
\titlerunning{PICS Project I. -- Metallicity Impact on PNLF}

\author{
    Lucas M. Valenzuela\inst{\ref{inst:usm}}
    \and
    Marcelo M. Miller Bertolami\inst{\ref{inst:ialp},\ref{inst:fcaglp}}
    \and
    Rhea-Silvia Remus\inst{\ref{inst:usm}}
    \and
    Roberto H. Méndez\inst{\ref{inst:ifa}}
}
\authorrunning{L. M. Valenzuela et al.}

\institute{
    Universitäts-Sternwarte, Fakultät für Physik, Ludwig-Maximilians-Universität München, Scheinerstr. 1, 81679 München, Germany\label{inst:usm}\\
    \email{lval@usm.lmu.de}
    \and
    Instituto de Astrofísica de La Plata, UNLP-CONICET, La Plata, Paseo del Bosque s/n, B1900FWA, Argentina\label{inst:ialp}
    \and
    Facultad de Ciencias Astronómicas y Geofísicas, UNLP, La Plata, Paseo del Bosque s/n, B1900FWA, Argentina\label{inst:fcaglp}
    \and
    Institute for Astronomy, University of Hawaii, 2680 Woodlawn Drive, Honolulu, HI 96822, USA\label{inst:ifa}
}

\date{Received XX Month, 20XX / Accepted XX Month, 20XX}

\abstract
{
Planetary nebulae (PNe) and their luminosity function (PNLF) in galaxies have been used as a cosmic distance indicator for decades, yet a fundamental understanding is still lacking to explain the universality of the PNLF among different galaxies.
Models for the PNLF have so far generally assumed near-solar metallicities and employed simplified stellar populations.
}
{
In this work, we investigate how metallicity and helium abundances affect the resulting PNe and PNLF, as well as the importance of the initial-to-final mass relation (IFMR) and circumnebular extinction, to resolve the tension between PNLF observations and previous models.
}
{
We introduce PICS (PNe In Cosmological Simulations), a PN model framework that takes into account the stellar metallicity and is applicable to realistic stellar populations obtained from both cosmological simulations and observations.
The framework combines current stellar evolution models with post-AGB tracks and PN models to obtain the PNe from the parent stellar population.
}
{
We find that the metallicity plays an important role for the resulting PNe: old metal-rich populations can harbor much brighter PNe than old metal-poor populations.
In addition, we show that the helium abundance is a vital ingredient at high metallicities and explore the impact on the PNLF of a possible saturation of the helium content at higher metallicities.
We present PNLF grids for different stellar ages and metallicities, where the observed PNLF bright end can be reached even for old stellar populations of \SI{10}{\giga\year} at high metallicities.
Finally, we find that the PNLFs of old stellar populations are extremely sensitive to the IFMR, potentially allowing for the production of bright PNe.
}
{
With PICS, we have laid the groundwork for studying how models and assumptions relevant to PNe affect the PNe and PNLF.
Two of the central ingredients for this are the metallicity and helium abundance.
Future applications of PICS include self-consistent modeling of PNe in a cosmological framework to explain the origin of the universality of the PNLF bright-end cutoff and use it as a diagnostic tool for galaxy formation.
}


\maketitle
%
%-------------------------------------------------------------------

\section{Introduction}
\label{sec:introduction}

For more than 30~years, planetary nebulae (PNe) have been used as distance indicators in the cosmic distance ladder through the luminosity function of their \OIII{} $\lambda5007$ magnitude in extragalactic systems, referred to as the planetary nebula luminosity function (PNLF) \citep{jacoby89:pnlfI}. Because it was observed that the PNLF has a universal bright end cutoff magnitude independent of the galaxy morphology, it has been found to work well as a distance indicator \citep[e.g.,][]{ciardullo+89:pnlfII, jacoby+90:pnlfV, rekola+05}.
Despite the wide usage, the underlying physics of the universality of the PNLF has remained a mystery in particular due to the question how old elliptical galaxies should be able to host any PNe as bright as the ones observed at the magnitude cutoff \citep{ciardullo12}. The argument for this has been that old stars at the main-sequence turnoff are too low in mass to produce such high emission in \OIII{}.
The PNLF has been successfully used as a distance indicator for galaxies at distances of up to 10--\SI{15}{\mega\parsec}, but has been limited to within \SI{20}{\mega\parsec} because of the limitations of narrowband filter imaging \citep{roth+21:I}.

In the past ten years, there have been two significant advances on the modeling and observational sides. With respect to modeling, \citet{miller_bertolami16} computed updated post-asymptotic giant branch (AGB) tracks of potential PN central stars. Those tracks were the first major update to the formerly computed stellar tracks from over 30~years ago \citep[e.g.,][]{schoenberner83:tracks, vassiliadis&wood94:tracks, bloecker95:tracks}. These new tracks featured a much faster post-AGB evolution with timescales shorter by a factor of three to ten and with brighter post-AGB stars. These led to several PN models being developed in recent years with a particular focus on improving the understanding of the universal bright end of the PNLF. Overall, the brighter post-AGB evolution has slightly diminished the large discrepancy between the observed bright PNe in old stellar populations and the models \citep[e.g.,][]{gesicki+18, valenzuela+19}. Further models have been developed with alternative attempts of explaining the bright end of the PNLF, including accreting white dwarfs (WDs; e.g., \citealp{soker06, souropanis+23}), blue straggler stars \citep[e.g.,][]{ciardullo+05}, and WD mergers \citep[e.g.,][]{yao&quataert23}.

On the observational side, the integral field unit (IFU) measurements with the instrument MUSE have made it possible to obtain the full spectra for each individual pixel, thereby making the detection of PNe and the removal of contaminants possible without follow-up measurements in pencil beam observations.
This enables the study of PNe at larger distances, and complements previously existing methods like the PN Spectrograph (PN.S; \citealp{douglas+02:PNS}; see also the extended PN.S early type galaxy survey, ePN.S; \citealp{pulsoni+18:ePNS}). In turn, the PN.S has the advantage of a significantly larger field of view (\SI{100}{\arcmin\squared} vs.~\SI{1}{\arcmin\squared} for MUSE) and higher spectral resolution, which has enabled detailed dynamical studies of galaxies using PNe \citep[e.g.,][for face-on disks]{aniyan+18, aniyan+21}. Using MUSE, several works have analyzed extragalactic PN populations, getting better data for previously studied objects or investigating new systems \citep[e.g.,][]{kreckel+17, spriggs+20, roth+21:I, scheuermann+22, soemitro+23:IV, jacoby+24:II}. Furthermore, the high spatial resolution of MUSE is expected to make it possible to reach distances of up to \SI{40}{\mega\parsec}, more than twice as far compared to previous limits \citep{roth+21:I}.

To establish a theoretical foundation for the PNLF and the usefulness of its bright end as a distance indicator, numerous approaches have been taken throughout the years to explain the properties of the PNLF. Some authors have created models for the entire population of PNe based on mock stellar populations \citep[e.g.,][]{mendez+93, mendez&soffner97, schoenberner+07:IV, valenzuela+19}, others have employed analytic properties of the PNLF \citep[e.g.,][]{rodriguez_gonzalez+15}.
In contrast, the majority of studies have focused on the bright end specifically, that is the maximum brightnesses that are achievable by PNe of certain stellar populations \citep[e.g.,][]{dopita+92, marigo+04:II, gesicki+18, souropanis+23, yao&quataert23}.
One of the findings from theory and observations has been that there should be a slight dependence of the PNLF bright end on metallicity, where the bright end becomes fainter for very metal-poor systems \citep[e.g.,][]{ciardullo&jacoby92:pnlfVIII, dopita+92, ciardullo+02:pnlfXII, ciardullo10}. However, the theoretical expectation that older stellar populations would produce PNLFs with their bright end being significantly dimmer is not observed \citep[e.g.,][stating that the PNLF bright end should already be more than 4~magnitudes dimmer for ages of \SI{10}{\giga\year}]{marigo+04:II, ciardullo12}.

Not only the bright end of the PNLF has drawn attention, but also the shape of the faint part of the PNLF: the general shape was introduced in its analytical form by \citet{ciardullo+89:pnlfII} as an exponential cutoff function with a parameter describing the slope on the faint side based on theoretical derivations of \citet{henize&westerlund63} and \citet{jacoby80}. This slope has been predicted to encode clues about the star formation history \citep[e.g.,][]{ciardullo10, longobardi+13, hartke+17} and has been investigated in particular for the different substructures in the outskirts of \m{31} \citep{bhattacharya+21:III}. For this, \citet{bhattacharya+21:III} also used the approach of fitting a superposition of two modes of the PNLF following \citet{rodriguez_gonzalez+15}. A further interesting feature that has been observed in the faint part of the PNLF is a dip for galaxies with recent star formation, that is a lower abundance of PNe at an intermediate magnitude \citep[e.g.,][]{jacoby&de_marco02, reid&parker10:III, soemitro+23:IV}. Even some models have been able to find such a dip in the PNLF for younger stellar populations \citep[e.g.,][]{mendez+08, valenzuela+19}.

With the multitude of observations and modeling approaches, the continual lack of fundamental understanding concerning the PNLF suggests that some of the previously employed assumptions are insufficient to address the problem at whole.
Indeed there are two central assumptions that have been commonly made in PN models over the years that do not represent realistic stellar populations as they are found in actual galaxies:
First, models have mostly only taken into account stellar evolution at metallicities around the solar metallicity, which has been done with respect to their main sequence (MS) and post-MS lifetimes, the initial-to-final mass relation (IFMR), and the post-AGB evolution \citep[e.g.,][]{gesicki+18, valenzuela+19, souropanis+23}.
Furthermore, even stellar evolution models employ many recipes based on measurements of the Sun and assume abundances based on the Milky Way \citep[e.g.,][]{miller_bertolami16}. For example, the helium abundance has mostly been assumed to scale linearly with metallicity, though this has been put into question recently \citep[e.g.,][]{clontz+25}.
Second, the studies that have attempted to model the entire PN population have set up artificial mock stellar populations with the goal of mimicking selected real galaxies \citep[e.g.,][]{mendez&soffner97, valenzuela+19}, but again relying on observations in the Milky Way, specifically the WD mass distribution. These models therefore fall short of accurately representing the large diversity of galaxies and star formation histories, which can strongly vary depending on intrinsic factors and the cosmological environment.

To overcome these limitations of current PN models, we introduce PICS (PNe In Cosmological Simulations), a modular framework for modeling PNe within any given stellar population, taking into account properties such as mass, age, metallicity, and the initial mass function (IMF).
In the era of hydrodynamical cosmological simulations, a new approach is further possible: realistic stellar populations with self-consistent properties in the cosmological context can be directly obtained for a large variety of galaxies from such simulations.
With PICS we present the very first method for bringing PNe into cosmological simulations and simultaneously improve on previous PN models by fully considering the metallicity and specific abundances beyond solar values.
Finally, the modular nature of the framework allows for controlled testing of different parameters and individual models, which we will also demonstrate in this paper.

In \cref{sec:model}, we present the PICS model structure and the implementation of the fiducial model, including the metallicity dependence on the individual modules.
By combining some of the modules, we then show how the metallicity and helium abundances affect the stellar evolution relevant for PNe in \cref{sec:metallicity}.
In \cref{sec:ssp_pnlfs}, we present grids of PNLFs resulting from different pairs of ages and metallicities and from varying the IFMR -- these are the building blocks for any real PNLF.
Finally, we summarize the results and conclude in \cref{sec:conclusion}.

\section{Model}
\label{sec:model}

In the following, we describe the building blocks of the PICS model. Its target is to take a given single stellar population (SSP) parameterized by a set of properties and determine the contained PN population. The obtained PN properties in particular include the magnitude of the emitted \OIII{} line, as it is the primary intrinsic observable quantity of an extragalactic PN. This procedure will make it possible to apply PICS to both observed galaxy star formation histories discretized by SSPs and to the stellar particles extracted from cosmological simulations (Soemitro et al.\ in prep.; Valenzuela et al.\ in prep.). The end result is then a full population of PNe for an entire system like a galaxy or galaxy cluster.

The model is designed to be modular, meaning that the individual parts described in the following can be replaced with different underlying models, and the selection of models will also be continuously expanded in future versions. In this work, we present the fiducial model and some alternative relations for the IFMR and how they affect the PNLF.
The source code and more information on using the model are made available online.\footnote{\url{https://lucasvalenzuela.de/PICS/}}
A summary of the general model functionality is presented in the proceedings by \citet{valenzuela+24:iaus}. The correction for nebular metallicity and the inclusion of circumnebular extinction are presented here for the first time.

\subsection{Single stellar population properties}
\label{sec:ssp_properties}

An SSP represents the population of stars having formed from a gas cloud where all the member stars share the same underlying age and chemical composition. The distribution of stellar masses within the SSP is given by an IMF.
In its simplest form, the input SSP properties need to include its total mass, an IMF, and its age. The age is needed to determine the initial mass range of stars now residing within the PN phase after having left the AGB track. The IMF is used to determine the fraction of stars that are within that initial mass range, and the total mass sets the absolute number of those stars. In addition to these properties, further quantities significantly influence the processes relevant for stellar evolution and PNe: in particular, we show in this work that the stellar metallicity is a vital parameter that affects the duration of stellar evolution on the MS and post-MS. In the future, individual abundances like helium or $\alpha$-element abundances will also be added as quantities that are taken into account.

In this work, each SSP is parameterized with the following quantities:
\begin{itemize}
    \item IMF, for this work we only consider that by \citet{chabrier03:imf}
    \item total mass $M_\mathrm{tot}$ of the SSP
    \item age $t$, that is the time since the formation of the SSP stars
    \item metallicity $Z$, in this work quantified as the initial metal mass fraction: $Z = M_\mathrm{met,init} / M_\mathrm{tot,init}$. As a point of reference, the solar metallicity lies between $Z = 0.01$ and 0.02 \citep[e.g.,][]{asplund+09, caffau+11, von_steiger&zurbuchen16, asplund+21}.
\end{itemize}

\subsection{Stellar lifetime function}
\label{sec:lifetime}

In the context of PN modeling, we define the stellar lifetime as the time from the birth of a star until it leaves the AGB phase and enters the post-AGB phase, during which a star can become a PN. It therefore roughly corresponds to the combined MS and post-MS lifetimes from other works \citep[e.g.,][]{renzini81, renzini&buzzoni86, padovani&matteucci93}. The purpose of this module in PICS is to return the initial stellar mass of the stars that now are passing through the PN phase after the age $t$.
Due to the short-lived nature of the PN phase (assumed as \SI{30000}{\year} here), the time spent after the departure from the AGB phase is negligible for the initial mass of the star. Then, disregarding binary interactions, we can assume that all PNe in a single stellar population are descendants of stars with the same initial mass.

For this work we took the lifetimes of the stellar evolution simulations from \citet{miller_bertolami16}, given in their tab.~2. For each of their considered metallicities ($Z = 0.0001$, 0.001, 0.01, and 0.02) and initial masses (ranging from $M_\mathrm{init} = 0.8$--\SI{4.0}{\Msun}, depending on the metallicity), we added up all the stated lifetimes: MS, red giant branch (RGB), core He-burning, early AGB, and thermally pulsing AGB (TP-AGB) phases as an M-type and carbon star.
Taking these lifetimes and the initial masses from \citet{miller_bertolami16}, we fit a double power law for each of the four metallicities ($Z = 0.0001$, 0.001, 0.01, and 0.02) by fitting a power law to the lower masses up to and including \SI{1.5}{\Msun} and a power law to the higher masses starting at \SI{1.5}{\Msun} (i.e., with one value in common). \Cref{tab:lifetime_fits} shows the fitting parameters for the free parameters of
\begin{equation}
    t / \si{\giga\year} = a \times (M_\mathrm{init} / \si{\Msun})^k
\end{equation}
for the low and high masses at the four metallicities, which lead to very good descriptions of the original data points, even in logarithmic space.

\begin{table}
    \centering
    \caption{The lifetime function double power law fit parameters for the low (${\leq}\SI{1.5}{\Msun}$) and high initial masses (${\geq}\SI{1.5}{\Msun}$) at the given metallicities $Z$ from \citet{miller_bertolami16}, parameterized as $t / \si{\giga\year} = a \times (M_\mathrm{init} / \si{\Msun})^k$.}
    \label{tab:lifetime_fits}
    \begin{tabular}{SSSSS}
        \hline
        {$Z$} & {$a_\mathrm{low}$} & {$k_\mathrm{low}$} & {$a_\mathrm{high}$} & {$k_\mathrm{high}$} \\
        \hline
        0.0001 & 11.80 & -3.58 & 7.98 & -2.55 \\
        0.001 & 9.72 & -3.52 & 6.54 & -2.45 \\
        0.01 & 6.41 & -3.45 & 4.64 & -2.45 \\
        0.02 & 5.95 & -3.40 & 4.75 & -2.65 \\
        \hline
    \end{tabular}
\end{table}

\begin{figure}
    \centering
    \includegraphics[width=\columnwidth]{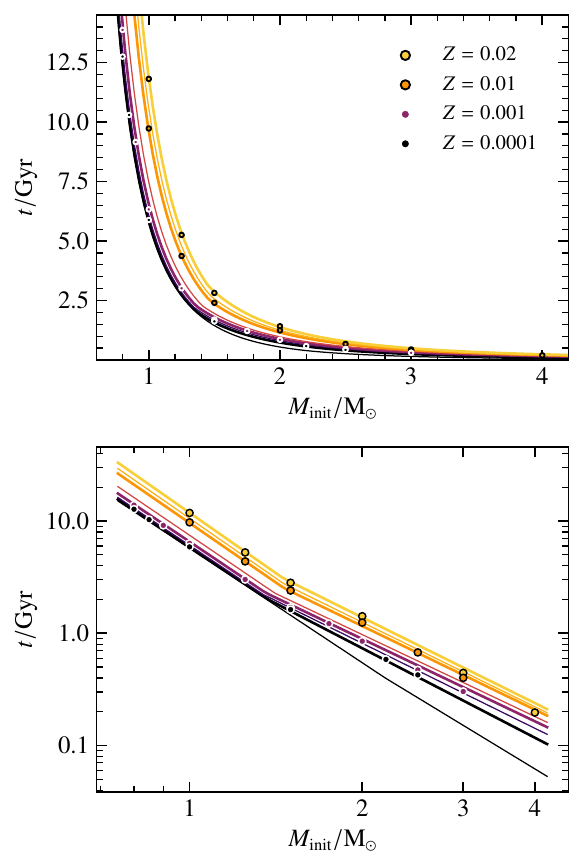}
    \caption{Lifetime function of stars until they reach the post-AGB phase as a function of their initial mass. \textit{Top}: Linear axes. \textit{Bottom}: Logarithmic axes, which makes it easier to distinguish the individual lines and data points. The points are from the models by \citet{miller_bertolami16}, and the lines are the fit relations to those data points. The thick lines are at the same metallicities as the ones from the original models ($Z = 0.0001$, 0.001, 0.01, and 0.02) and the thin lines are at $Z = 0.00001$, 0.0003, 0.003, and 0.014, where the latter three values are the logarithmic midpoints of the original metallicities.}
    \label{fig:lifetime}
\end{figure}

These fits only correspond to the four discrete metallicities from \citet{miller_bertolami16}. To obtain a continuous representation of the lifetime function across metallicities, we found the following relations to describe the change of the four double power law parameters with metallicity\footnote{See the implementation in the source code on GitHub for the exact values with more significant digits.}:
\begin{align}
    a_\mathrm{low} &= 78.27 \times Z^{0.649} + 5.67, \label{eq:alow} \\
    a_\mathrm{high} &= 107.5 \times Z^{0.880} + 4.58, \label{eq:ahigh} \\
    k_\mathrm{low} &= -0.943 \times Z^{0.400} - 3.38, \label{eq:klow} \\
    k_\mathrm{high} &= -3.29 - 0.654 \times (\log Z) - 0.124 \times (\log Z)^2 \label{eq:khigh}.
\end{align}
The lifetime is then determined as the maximum between $t_\mathrm{low}$ and $t_\mathrm{high}$ computed from the respective power law parameters obtained from the stated relations. The lifetime function is shown for the original and some selected further metallicities in \cref{fig:lifetime}, showing the continuous nature of the here derived expression.
In \cref{sec:metallicity}, we present how the extrapolation of this relation to higher metallicities compares to detailed evolutionary models computed at such higher metallicities.

The obtained lifetime functions at the four original metallicities (\cref{eq:alow,eq:ahigh,eq:klow,eq:khigh}) as well as at intermediate and one lower metallicity ($Z=0.00001$) are shown in \cref{fig:lifetime} as lines, colored by metallicity. The circular points indicate the values obtained from tab.~2 of \citet{miller_bertolami16} and lie directly on the respective thick solid functional lines. The thin lines, which are at interpolated and lower metallicities, lie between the thick lines, just as they should. The power law slope $k_\mathrm{high}$ becomes steeper with decreasing metallicity, such that the extrapolation to $Z=0.00001$ leads to a very small change in slope towards larger masses. We therefore conclude that our determined analytical lifetime function is suitable for metallicities of at least $Z=0.00001$--0.02.

While other PN studies have generally implicitly or explicitly assumed a fixed metallicity for their underlying mock stellar populations, oftentimes $Z = 0.01$ or $Z = 0.02$ \citep[e.g.,][]{gesicki+18, souropanis+23, valenzuela+19}, \cref{fig:lifetime} shows that the metallicity plays a very significant role for stars of any age. For a fixed stellar age, varying the metallicity from low values of $Z = 0.0001$ to $Z = 0.02$ results in a change of more than $M_\mathrm{init} = \SI{0.2}{\Msun}$. In contrast, large absolute changes of stellar age for $t \gtrsim \SI{5}{\giga\year}$ only slightly affect the initial mass, which becomes relevant towards lower masses as the mass approaches the minimum viable mass for PNe to form, however.
For old elliptical galaxies the metallicities tend to be high \citep[e.g.,][]{li+18:gradients}, which means that especially for those types of galaxies it is important to take into account the actual underlying stellar metallicity. The implications of a higher metallicity are the following: at a fixed stellar age, more metal-rich stars now passing through the PN phase are more massive than metal-poor stars that are PNe because more metal-rich stars evolve more slowly at a fixed stellar mass. This leads to more massive central stars with higher temperatures and luminosities, thus giving rise to brighter PNe, which is a possible way of populating the bright side of the PNLF with PNe from older stellar populations.

\subsection{Initial-to-final mass relation}
\label{sec:ifmr}

Having obtained the initial mass of stars that are now PNe, the IFMR is next used to determine the final mass based on the initial mass and metallicity of the star, that is how massive the central star of the PN is.
This relation has been mainly investigated through WD stars in star clusters in the solar neighborhood \citep[e.g.,][]{cummings+18}, though it has also been estimated from composite stellar populations \citep[e.g.,][]{el_badry+18}.

\begin{figure}
    \centering
    \includegraphics[width=\columnwidth]{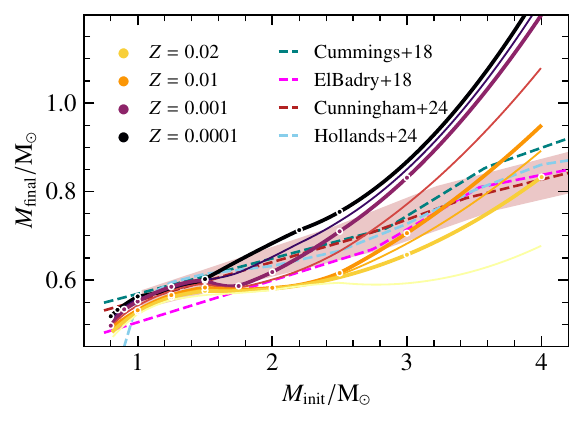}
    \caption{Initial-to-final mass relation of stars. The points are from the models by \citet{miller_bertolami16}, and the lines in the corresponding colors are the fit relations to those data points. The thick lines are at the same metallicities as the ones from the original models ($Z = 0.0001$, 0.001, 0.01, and 0.02) and the thin lines are at $Z = 0.00001$, 0.0003, 0.003, 0.014, 0.05, where the middle three values are the logarithmic midpoints of the original metallicities.
    Additionally, four observationally based IFMRs from nearby WD measurements at solar metallicity are shown from \citet{cummings+18}, \citet{el_badry+18}, \citet{cunningham+24}, and \citet{hollands+24}, as well as the $1\sigma$ uncertainty band of \citet{cunningham+24}.}
    \label{fig:ifmr}
\end{figure}

For the fiducial model, we create a metallicity-dependent IFMR based on the values obtained from the models of \citet{miller_bertolami16}, which includes the four metallicities $Z = 0.0001$, 0.001, 0.01, and 0.02, just as for the lifetime function (\cref{sec:lifetime}). To sample more data points between the simulated pairs of initial and final mass values at each metallicity, a forward-looking quadratic interpolation \citep{bhagavan+24:datainterpolations.jl} is used. Then, a 2D linear interpolation is applied between the four logarithmic metallicities to the more finely sampled values, which is the resulting metallicity-dependent IFMR. This IFMR can be applied in two extrapolation modes: for metallicities outside of the range of $Z = 0.0001$--0.02, the values at the respective bounding metallicity can be used as a constant beyond that value, or the linear fit can be extrapolated beyond the original metallicity range. In \cref{fig:ifmr}, we show the resulting IFMR with extrapolated values at the same metallicities as shown for the lifetime function in \cref{fig:lifetime}, plus the higher metallicity of $Z = 0.05$. The colored dots show the original data points from table~2 of \citet{miller_bertolami16}, between which the interpolation took place.

The dependence on metallicity of the IFMR shows that more metal-rich stars at a fixed initial mass end up having lower final masses than metal-poor stars. As more massive stars are hotter and more luminous in the post-AGB phase and thus lead to brighter PNe, this counteracts the effect that metallicity has on the lifetime function, which leads to stars with higher initial masses entering the PN phase.

It is worth noting that the metallicity dependence of the IFMR of the models of \citet{miller_bertolami16} is a consequence of the adopted winds. The metallicity dependence of cold winds is still poorly understood, in particular in the first RGB phase. The resulting theoretical IFMRs should therefore be taken as educated guesses. Most reliable IFMRs come from semi-empirical studies of star clusters in the solar neighborhood, which are only available for near-solar metallicities, however.

As the IFMR in general and also its metallicity dependence are still poorly understood, it is important to better understand the dependence of the resulting PNe in a given SSP on the used IFMR. For this, we additionally consider four IFMRs derived from observed WDs in the Milky Way: \citet{cummings+18} determined their IFMR from 80~WDs located in 13~star clusters, \citet{el_badry+18} estimated the IFMR from Gaia measurements of over 1000~WDs within \SI{100}{\parsec} fo the Sun, \citet{cunningham+24} studied 40~WDs with spectroscopic measurements from Gaia data within \SI{40}{\parsec} of the Sun, and \citet{hollands+24} studied 90~WD binaries from Gaia data with distances of up to \SI{250}{\parsec}. All four of these IFMRs correspond to near-solar metallicities. We also show IFMRs from the four studies in \cref{fig:ifmr}, where we use the adopted MIST-based IFMR of \citet{cummings+18} and linearly extrapolate the IFMRs down to lower initial masses, as \citet{cunningham+24} and \citet{hollands+24} only fit their IFMRs down to \SI{1}{\Msun}, and \citet{cummings+18} to \SI{0.83}{\Msun}.
The uncertainties of these IFMRs are on the order of $\sigma_{M_\mathrm{final}} \approx \SI{0.025}{\Msun}$, for which we also show the uncertainty region of \citet{cunningham+24}. Further, the systematic uncertainties are on the order of $\SI{0.1}{\Msun}$, as seen by using different underlying stellar evolution model is employed \citep{cummings+18}.
Only the low-mass IFMR from \citet{el_badry+18} produces significantly lower final masses than the other relations. According to \citet{cunningham+24}, the reason for this is that \citet{el_badry+18} use a different method for obtaining the minimum WD mass that can be created through single stellar evolution.
While these relations are all monotonic, we note that it is still debated whether this is the case or if there is a dip around $M_\mathrm{init} \approx 1.5$--\SI{2}{\Msun} \citep[e.g.,][]{marigo+20} due to the large uncertainties.

Overall, there is a general agreement on what the IFMR looks like within the measurement uncertainties and their large systematics, where the theoretical metallicity-dependent IFMR from \citet{miller_bertolami16} predicts overall lower final masses at solar metallicity than the observationally based IFMRs, especially at final masses of $M_\mathrm{init} \approx 1.5$--\SI{3}{\Msun} around the potential dip. Still, the small differences in final mass can lead to significant changes in the resulting PN properties, especially moving towards final masses below \SI{0.6}{\Msun}. In \cref{sec:ssp_pnlfs}, we investigate how the different IFMRs impact the resulting PN population with respect to their PNLF.

\subsection{Post-AGB stellar evolution tracks}
\label{sec:postagb_tracks}

With the final mass of the star in question, the evolving properties of the central star can be obtained from post-AGB stellar evolution tracks. In particular, the luminosity $L$ and effective temperature $T_\mathrm{eff}$ can be directly obtained with the tracks in the Hertzsprung-Russell diagram, given the final mass and an identifying time. For PICS, we assume the lifetime of a PN can be up to \SI{30000}{\year} \citep{valenzuela+19}. For a given central star, the time since leaving the post-AGB phase is therefore randomly drawn from a uniform distribution between 0~and \SI{30000}{\year}, as the timescale is negligible compared to the timescale of the MS and post-MS.

\begin{figure}
    \centering
    \includegraphics[width=\columnwidth]{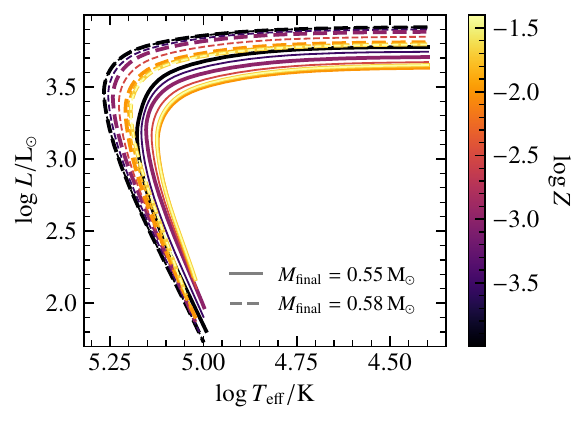}
    \caption{Post-AGB tracks in the Hertzsprung-Russell diagram of stars at two final masses $M_\mathrm{final} = \SI{0.55}{\Msun}$ and \SI{0.58}{\Msun} for a range of metallicities. The tracks are interpolated between the data tables from \citet{miller_bertolami16}. The metallicity colorbar corresponds to the same coloring as found in \cref{fig:lifetime,fig:ifmr}. The thick lines are at the same metallicities as the ones from the original models ($Z = 0.0001$, 0.001, 0.01, and 0.02) and the thin lines are at $Z = 0.00001$, 0.0003, 0.003, 0.014, 0.05, where the middle three values are the logarithmic midpoints of the original metallicities.}
    \label{fig:postagb_tracks}
\end{figure}

In this work, we use the post-AGB tracks of the H-burning stars from \citet{miller_bertolami16} and interpolate them to obtain the central star luminosity and effective temperature as a function of metallicity, final mass, and the post-AGB age (see \appref{app:postagb_tracks} for details on the interpolation). Following the approach of \citet{valenzuela+19}, we take the post-AGB age to be the time since the temperature reached a value of $T_\mathrm{eff} = \SI{25000}{\kelvin}$ ($\log T_\mathrm{eff} / \si{\kelvin} \approx 4.40$).
The interpolated post-AGB tracks at two final masses of $M_\mathrm{final} = 0.55$ and \SI{0.58}{\Msun} (of which only the former mass was explicitly computed by \citealp{miller_bertolami16}) for a range of different metallicities can be seen in \cref{fig:postagb_tracks}. Clearly, the interpolations successfully produce intermedial values, in particular between the originally simulated metallicities.
For metallicities outside of the range between $Z=0.0001$ and 0.02, we simply use the respective tracks at $Z=0.0001$ or $Z=0.02$.

Additionally, it can here be seen that more metal-poor stars at a fixed final mass have a higher luminosity and can reach higher temperatures than metal-rich stars, as was also shown by \citet{miller_bertolami16}. Like the metallicity-dependent IFMR from \citet{miller_bertolami16} presented in \cref{sec:ifmr}, this counteracts the effect the lifetime function has that one would expect more metal-rich stars to host brighter PNe, as well as the lower number of oxygen atoms in the nebula, which is also expected to limit the brightness of the \OIII{} line..
The dependence of the maximum luminosity and effective temperature of a star on its metallicity and age is therefore not trivial.

\subsection{Planetary nebula model}
\label{sec:pn_model}

The central star properties can finally be used to infer the PN emissions. Particularly, the \OIII{} intensity is of great importance for studies of extragalactic PNe. Additionally, properties such as the \Hbeta{} intensity may be of further relevance and has been studied for extragalactic PN populations as well \citep[e.g.,][]{reid&parker10:III}.

In this work we use the PN model from \citet{valenzuela+19}. To our knowledge it is the only current PN model that not only addresses the brightest PNe at the bright end of the PNLF, but can also be used to produce dim PNe that are consistent with observations, while being applicable to large numbers of central stars without exceedingly high computational cost. The model computes the maximum total intensity of \Hbeta{} emission assuming optical thickness, $I(\Hbeta{})$, based on the luminosity of the central star luminosity and its blackbody radiation given by the effective temperature. Next, the ratios between $I(\Hbeta{})$ and the intensity of the \OIII{} emission line $I(5007)$ that have been observed for PNe in the Milky Way and the LMC are used to empirically obtain $I(5007)$ for a given PN. This ratio is drawn from a Gaussian distribution and is modified by some physically motivated prescriptions based on stellar mass and temperature. We give an overview of this prescription in \cref{app:intensity_ratio}.

\begin{figure*}
    \centering
    \includegraphics[width=\textwidth]{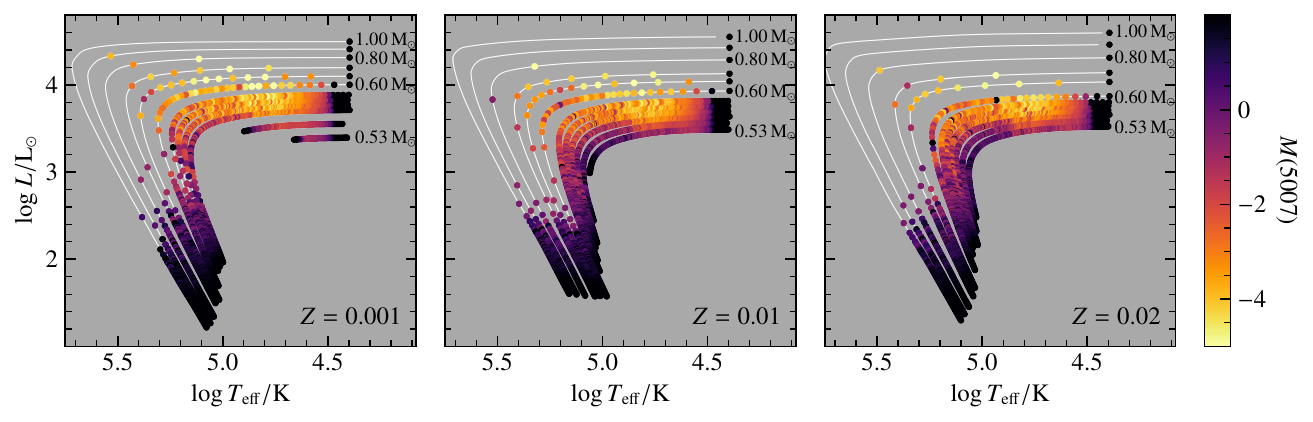}
    \caption{Post-AGB tracks in the Hertzsprung-Russell diagram of stars at metallicities $Z=0.001$, $0.01$, and $0.02$ (left, middle, and right, respectively), colored by the $\lambda5007$ magnitude. The tracks are shown for stars at different final masses $M_\mathrm{final} = 0.53$, 0.54, 0.55, 0.56, 0.57, 0.58, 0.60, 0.65, 0.70, 0.80, 0.90, and \SI{1.00}{\Msun}. For each final mass, stars at 600 evenly spaced post-AGB ages between 0 and \SI{30000}{\year} are plotted. The white lines indicate the tracks for the selected central star masses.}
    \label{fig:postagb_tracks_m5007}
\end{figure*}

Finally, an absorbing factor $\mu$ is applied to the intensities to account for ionizing stellar photons that are not absorbed by the nebula. The absorbing factor is equivalent to the fraction of ionizing photons that are actually absorbed and is simply multiplied with the previously obtained intensity to get the observed value of $I(5007)$. The absorbing factor for a given star is drawn randomly because of possible asymmetries in the nebula that can lead to a leakage of ionizing photons (for more details see \citealp{mendez&soffner97, valenzuela+19}). Additionally, its maximally allowed value is decreased over the lifetime of the PN, which accounts for the nebula dissipating over time. The prescription for the absorbing factor is described in \cref{app:absorption_factor}.
One aspect that is not included in the model of \citet{valenzuela+19} is the internal dust extinction, which becomes more important for younger stars \citep[e.g.,][]{jacoby&ciardullo25}, which means in general for higher stellar masses.
We add this effect as a further step in the PICS framework in \cref{sec:circumnebular_extinction}.

In \cref{fig:postagb_tracks_m5007} we show the post-AGB tracks in the Hertzsprung-Russell diagram of stars with certain final central star masses at metallicities of $Z = 0.001$, 0.01, and 0.02. The data points show individual PNe at evenly spaced ages between 0 and \SI{30000}{\year} starting from the given track reaching $T_\mathrm{eff} = \SI{25000}{\kelvin}$. The PNe are colored according to their \OIII{} $\lambda5007$ absolute magnitude, $M(5007) = -2.5 \log I(5007) - 13.74$ with $I(5007)$ being the intensity seen \SI{10}{\parsec} away, as defined by \citet{jacoby89:pnlfI}. According to the model, the brightest PNe for a given central star mass are located on the rising temperature track, after which the intensity of the \OIII{} line decreases again due to the dissipation of the nebula over time and the eventual drop of the star's luminosity. Here it also becomes apparent how quickly more massive stars ($M_\mathrm{final} \gtrsim \SI{0.6}{\Msun}$) evolve along the tracks compared to less massive stars, for example at $Z=0.001$: two data points are separated by \SI{50}{\year}, which means that after only a little more than \SI{100}{\year}, a central star of mass \SI{0.8}{\Msun} reaches its peak temperature. As a result, even though more massive stars have the potential of developing brighter PNe, the likelihood of observing such a PN in its bright stage is much lower for more massive stars due to their fast evolution.

The PN model by \citet{valenzuela+19} was originally developed for use with the post-AGB tracks from \citet{miller_bertolami16} at a metallicity of $Z = 0.01$ and the \OIII{} to \Hbeta{} intensity ratio was empirically based on Milky Way and LMC PNe, which span a range of metallicities. As there is no obvious step in the model's recipes where it would be trivial to add a direct dependence on metallicity, in this work we apply the PN model to central stars of all metallicities without modification.
While the model therefore statistically works well enough for composite populations of PNe that roughly have the same metallicity distribution as the Milky Way and LMC PNe, the intrinsic dependence on metallicity is lost, where lower abundances of oxygen should generally lead to less \OIII{} emission. For this reason, only applying the PN model by \citet{valenzuela+19} to especially metal-poor PNe will result in overluminous PNe.

Multiple studies have investigated the dependence of the maximum \OIII{} flux on the metallicity. This relation is not straightforward as \OIII{} is also one of the primary coolants of the nebula \citep[e.g.,][]{jacoby89:pnlfI}. Using nebular models, \citet{jacoby89:pnlfI} found that small changes in the metallicity at solar-like abundances resulted in variations of the \OIII{} flux by the square root.
Later models by \citet{dopita+92} and \citet{schoenberner+10} showed a quadratic form in logarithmic space around solar abundances for the peak luminosity of a given PN (see fig.~27 from \citealp{schoenberner+10}).
In this work we consider all three possible relations as direct corrections to the obtained \OIII{} fluxes, extrapolating them beyond solar abundances to both low and high metallicities. As the nebular model by \citet{valenzuela+19} was not based on a particular metallicity to which a simple nebular metallicity correction can be applied, we needed to estimate a base metallicity $Z_\mathrm{base}$ of the model from which it is possible to use the abundance differences to determine the correction required for the \OIII{} flux. Because Milky Way and LMC PNe were used for creating the empirical model by \citet{valenzuela+19}, the value has to be equal to or lower than the maximum metallicities of both galaxies. The overall more metal-poor LMC has an upper metallicity of around $Z = 0.008$ \citep{harris&zaritsky09}, which is consistent with the upper oxygen abundance found for its PNe of $12 + \log(\mathrm{O/}\mathrm{H}) = 8.5$--8.6 \citep{dopita&meatheringham91:II, richer93}.
At the same time the base metallicity value may not be too low as the observed nebulae have to have a sufficient amount of oxygen available to become PNe in the first place.
We therefore estimate this base metallicity as $Z_\mathrm{base} = 0.007$, a value near the upper metallicity value of the LMC and also roughly half of the solar metallicity. We decided on this value because it is within the reasonable metallicity range of both the Milky Way and the LMC. However, it is important to note that this is not a physically derived value due to the nature of the empirical model being based on a distribution of metallicities.

For the square root relation of \citet{jacoby89:pnlfI}, we obtain a nebular metallicity correction factor of
\begin{equation}
    p_{5007,\mathrm{J89}}(Z) = \left( \frac{Z}{Z_\mathrm{base}} \right)^{0.5}.
\end{equation}
For \citet{dopita+92}, we assumed a linear conversion between the metallicity and oxygen abundance [O/H], so with our choice of the base metallicity we obtain a correction factor of
\begin{align}
    & p_{5007,\mathrm{D92}} = 10^{\log F_{5007,\mathrm{D92}}(\log Z / Z_\odot) - \log F_{5007,\mathrm{D92}}(\log Z_\mathrm{base} / Z_\odot)}, \\
    & \log F_{5007,\mathrm{D92}}([Z]) = -0.8791 + 0.1459 [Z] - 0.3013 [Z]^2,
\end{align}
where the latter equation is taken from their eq.~4.1 with $[Z] = \log Z / Z_\odot$ and we take $Z_\odot = 0.0134$ from \citet{asplund+09}, which is also the solar metallicity that \citet{schoenberner+10} assumed in their models.
Finally, for \citet{schoenberner+10} we fit two quadratic functions to their data points for their two final masses and used the mean of each of the fit parameters to obtain the following correction factor:
\begin{align}
    & p_{5007,\mathrm{S10}} = 10^{\log F_{5007,\mathrm{S10}}(\log Z / Z_\odot) - \log F_{5007,\mathrm{S10}}(\log Z_\mathrm{base} / Z_\odot)}, \\
    & \log F_{5007,\mathrm{S10}}([Z]) = -0.24 ([Z] - 0.78)^2.
\end{align}
We show the direct comparison of the three correction factors as a function of metallicity in \cref{app:nebular_metallicity}.
As the correction factor only depends on metallicity, the values of $M(5007)$ in \cref{fig:postagb_tracks_m5007} are simply offset by a fixed value per panel when applying one of the three nebular metallicity corrections.

While the nebular metallicity relations from the three works were originally constructed to describe how the fluxes of the brightest PNe vary with metallicity, we have decided to use these relations for correcting the \OIII{} fluxes of all PNe irrespective of their properties.
We also note the caveat of applying the nebular metallicity correction to the resulting fluxes from the model by \citet{valenzuela+19}: as the model was tuned to PNe with a range of metallicities, having to pick a single base metallicity from which the correction is determined leads to an excessive brightness of PNe on the slightly metal-rich side and an excessively strong dimming on the slightly metal-poor side.
As the disadvantages of not applying such a correction outweigh this caveat, however, we pick as the fiducial model the nebular metallicity correction by \citet{dopita+92}, which does not boost the \OIII{} flux as much as the other two relations (see \cref{fig:nebular_metallicity}).

Finally, we comment on the fact that for our current fiducial nebular model, we only take into account the general metallicity mass fraction $Z$ and not any specific abundances. On the one hand, this has the big advantage of keeping the model simple, especially in light of the large parameter space already spanned through the stellar evolution models, which has also been the focus for this paper. With the aim of applying PICS to cosmological simulations, most chemical evolution models are not yet able to find good agreement with observed abundances, whereas the total metallicity mass fractions are in general consistent with observations.
On the other hand, the simple linear scaling assumptions of different abundances like helium, oxygen, and iron with $Z$ are certainly not entirely correct and may lead to systematics and biases in the modeled PN emission lines. For instance, accounting for alpha elements independently of iron-peak elements may be relevant when treating galaxies with different star formation histories like typical elliptical versus disk galaxies. Similarly, increased helium abundances could affect the number of ionizing photons available to ionize hydrogen \citep[e.g.,][]{osterbrock&ferland06}.
As the focus of this paper is on the advances of stellar evolution modeling in the context of PNe, treating the nebular physics with a more detailed model will be the subject of future work, further improving the PICS model.

\subsection{Circumnebular extinction}
\label{sec:circumnebular_extinction}

During the AGB phase stars create and eject dust through their stellar winds, which may lead to extinction of the PN emission. As more massive stars evolve more quickly along the post-AGB tracks and thus reach high temperatures to ionize the nebula shortly after the AGB phase, these PNe would suffer a larger extinction than PNe of low-mass stars when they reach their brightest \OIII{} magnitude. \Citet{ciardullo&jacoby99} suggested that this effect could explain how the bright end of the PNLF is even universal for PNe in younger stellar populations, where the central stars of the PNe would be more massive. For the bulge of \m{31}, \citet{davis+18} found that a significant fraction of the brightest PNe are strongly extincted. Most recently, \citet{jacoby&ciardullo25} presented the observed relation between central star mass and circumnebular extinction measured for the brightest PNe in the LMC and \m{31}, which they fit through linear relations. Their work extends on \citet{ciardullo&jacoby99} and presents strong indications that this effect indeed could explain how younger galaxies have the same universal bright-end cutoff of the PNLF as the circumnebular extinction could bring the overly bright PNe down to the cutoff.

As PICS has the objective to produce PN properties for direct comparison with observations, accounting for circumnebular extinction is an important last step for obtaining the actually observed \OIII{} flux. While it is in part possible to determine the circumnebular extinction of PNe in the Local Group, this is impossible for most extragalactic systems.
The relations for extinction as a function of final mass provided by \citet{jacoby&ciardullo25} are only valid for the PNe in the brightest magnitude. As the dust becomes more diffuse over the lifetime of a PN, the extinction is expected to decrease. Under the assumption that the evolution of dust extinction is the same for all PNe, we derived the extinction relation as a function of post-AGB age that recovers the relation by \citet{jacoby&ciardullo25} for the PNe at metallicity $Z = 0.01$ as they reach their peak brightness. We chose the metallicity of $Z = 0.01$ as a reasonable value between the expected metallicities of the brightest PNe in \m{31} and the LMC. We note that using slightly lower or higher metallicities do not affect the resulting extinctions significantly.

We first determined the post-AGB age corresponding to the maximal brightness as a function of final mass at $Z = 0.01$ and fit the following broken linear relation to it (see \cref{app:circumnebular_extinction} for the details of the derivation):
\begin{equation}
    M_\mathrm{final,bright} = \max \left\{ \begin{aligned}
        1.014 - 0.1704 \log(t_\mathrm{post-AGB} / \si{\year}) \\
        0.729 - 0.0479 \log (t_\mathrm{post-AGB} / \si{\year})
    \end{aligned} \right\}
    \label{eq:brightest_ages}
\end{equation}
In this work we use the combined orthogonal regression fit to the bright LMC and \m{31} PNe of \citet[][Combined OR in their table~2]{jacoby&ciardullo25}, which takes the corresponding brightest final mass as argument (independent of what the actual final mass is):
\begin{equation}
    c_{\mathrm{H}\beta} = 7.05 M_\mathrm{final,bright} / \Msun - 3.97,
\end{equation}
where the extinction in H$\beta$, $c_{\mathrm{H}\beta}$, is used to obtain the \OIII{} extinction, $A_{5007}$, through the \citet{cardelli+89} extinction law with $A_V = R_V E(B-V)$, $R_V = 3.1$, and $E(B-V) = c_{\mathrm{H}\beta} / 1.47$ \citep{ciardullo&jacoby99}:
\begin{equation}
    A_{5007} = \frac{c_{\mathrm{H}\beta}}{1.47} \left( R_V a(x) + b(x) \right),
\end{equation}
where $a(x)$ and $b(x)$ are defined for the optical/near-infrared range in eqs.~3a and 3b from \citet{cardelli+89} with $x = \SI{1}{\micro\meter} / \SI{5007}{\angstrom}$.
In this way, the extinction only depends on the post-AGB age of a PN.
While in this work we only apply the one circumnebular extinction relation from \citet{jacoby&ciardullo25}, we will investigate the effects of variations of the relation in a forthcoming paper (Valenzuela et al.\ in prep.).

\subsection{Number of planetary nebulae}
\label{sec:pn_number}

The previous sections dealt with the modeling of a single PN based on the IMF, age and metallicity of the SSP. How many PN this SSP produces is additionally dependent on the total mass. For this the IMF is integrated between the initial mass obtained from the SSP age (\cref{sec:lifetime}) and the initial mass obtained from the SSP age minus \SI{30000}{\year} because those are the stars that are still in the PN phase. The integrated value is then multiplied with the total mass to obtain the expected total number of PNe in that SSP. The actual integer number is then determined by rounding down or up randomly based on the decimal places, for example an expected value of 10.2 would lead to 10~PNe being selected with an \SI{80}{\percent} chance and 11~PNe with a \SI{20}{\percent} chance.

The number of PNe is obtained after the initial stellar mass has been found from the lifetime function. The final mass and the PN properties are then determined individually for each of those PNe since values such as the post-AGB age or the absorbing factor can be different for PNe even within the same SSP.

\section{The effect of metallicity and helium abundances}
\label{sec:metallicity}

\subsection{Metallicity}

In the previous section it can be seen that multiple aspects of the emergence of PNe depend on the metallicity: For a given initial mass, more metal-rich stars remain on the MS and post-MS for a longer period of time (\cref{fig:lifetime}), and they lose more mass assuming the IFMR from \citet{miller_bertolami16}, leading to smaller final masses after the AGB phase (\cref{fig:ifmr}). For the post-AGB tracks at a given final mass, more metal-rich stars have lower luminosities and reach lower maximum temperatures before entering the cooling track (\cref{fig:postagb_tracks}). Finally, while not depicted in this work, stars of a given final mass move along the Hertzsprung-Russell diagram at roughly the same speed, that is they heat up and cool down on a similar timescale (fig.~8 of \citealp{miller_bertolami16}).

\begin{figure*}
    \centering
    \includegraphics[width=\textwidth]{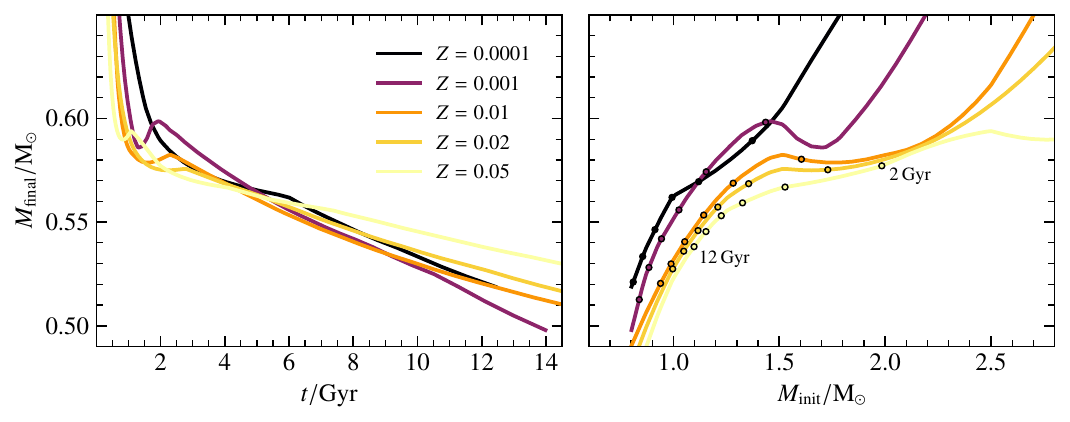}
    \caption{Final masses as a function of stellar lifetime (left) and of initial mass (right) for selected metallicities ($Z = 0.0001$, 0.001, 0.01, 0.02, and 0.05) based on the models of \citet{miller_bertolami16}.
    The final mass y-axis range is reduced to the primarily relevant values in the regime of older stellar populations of $t > \SI{2}{\giga\year}$.
    For the initial-to-final mass relation, the masses reaching the post-AGB phase after $t=2$, 4, 6, 8, 10, and \SI{12}{\giga\year} are marked with circles.
    }
    \label{fig:metallicity_effect}
\end{figure*}

It is not directly clear from the individual relations themselves how the metallicity affects the central stars of the PNe when combining the lifetime function and the IFMR. To shed more light on this, we show how the final stellar mass depends on the age with varying metallicity in the left panel of \cref{fig:metallicity_effect}. For a large age range, most of the metallicities lead to similar final masses: at ages of 2--\SI{7}{\giga\year} all but $Z=0.001$ lie closely together, and at ages of 5--\SI{11}{\giga\year}, all but $Z=0.05$ lie closely together. Based on this model, this implies that there is a nearly universal final mass relation with age independent of metallicity for these ages, with variations of around 0.01--\SI{0.02}{\Msun} at a given age. For lower and higher ages, however, the final masses begin to deviate more from one another. For stellar populations less than \SI{2}{\giga\year} old, stars becoming PNe that are more metal-rich tend to have smaller final masses. For ages above 8--\SI{10}{\giga\year} and with $Z > 0.001$, the trend emerges under the current model assumptions that more metal-rich stellar populations produce PNe with larger final masses due to their longer lifetimes compared to their lower-mass counterparts. While these trends are not monotonic in the shown metallicity range, old stellar populations that are especially metal-rich above solar metallicity produce significantly more massive central stars of PNe than at lower metallicities, still reaching final masses of \SI{0.55}{\Msun} after \SI{9}{\giga\year}.

Based on these relations, we can now highlight the importance of taking the metallicity into account for especially metal-rich systems: if one were to follow the approach of some previous studies of assuming a fixed roughly solar-like metallicity of $Z=0.01$ \citep[e.g.,][]{gesicki+18, valenzuela+19}, then for ages above \SI{5}{\giga\year} the final masses at that fixed metallicity would be systematically lower than if taking the correct higher metallicity into account (see the left panel of \cref{fig:metallicity_effect}). The information can also be presented in the IFMR plot from \cref{sec:ifmr} by marking specific times along the IFMRs at each metallicity. The right panel of \cref{fig:metallicity_effect} shows the IFMR for a more limited range of initial and final masses than shown in \cref{fig:ifmr} where the points along the IFMRs are marked at equally spaced lifetimes between 2 and \SI{12}{\giga\year}. Especially at higher metallicities it can be seen that very old stellar populations with $t > \SI{6}{\giga\year}$ have stars of higher initial and final masses reach the PN phase, resulting in old and metal-rich elliptical galaxies having more massive central stars of PNe than the old stellar population of a more metal-poor spiral galaxy, for example.

One caveat of the above conclusions based on the highest shown metallicity $Z=0.05$ is that this metallicity lies outside the range originally modeled by \citet{miller_bertolami16}.
Since those models only reached a metallicity of $Z=0.02$, it cannot simply be assumed that the extrapolation of the lifetime function to higher metallicities is applicable using the fit analytic relation (\cref{sec:lifetime}).
One difficulty in this matter is that stellar evolution and in particular the AGB phase are still poorly understood for super-solar metallicity stars. This is largely due to such metal-rich stars being mostly observed only in the Galactic bulge or in \m{31} for massive stars \citep[e.g.,][]{pietrinferni+13, mcdonald+22}. Some studies have run models for super-solar metallicity stars, including the PARSEC \citep{bressan+12:parsec} and BaSTI \citep{pietrinferni+13} stellar evolution models, additionally \citet{althaus+09} investigated potential high-metallicity progenitors of WDs, and \citet{nanni+14} ran super-solar metallicity models to study the dust production in pulsing AGB stars.
As stated by \citet{pietrinferni+13}, there are differences between their BaSTI and the PARSEC models in part because of different initial helium abundance assumed in the models.

\subsection{Helium abundance}

Not only the initial helium abundance is uncertain: its scaling with the overall metallicity has been primarily measured at lower metallicities in the Milky Way and in \HII{} regions. Mostly, studies have quantified a slope $\Delta Y / \Delta Z$ as the cosmic production rate of helium, which has been estimated to be between 1 and~3 \citep[e.g.,][]{jimenez+03, izotov&thuan04, casagrande+07}. However, not all authors assume a linear slope, especially at the lowest metallicities \citep[e.g.][]{keszthelyi+24}. While \citet{miller_bertolami16} assumed a slope of 2 for the initial abundances of the stellar evolution models, it is unclear how exactly the helium-to-metallicity relation continues beyond the upper metallicity of $Z=0.02$. \Citet{casagrande+07} even suggest that the slope becomes smaller towards higher metallicities, though the large uncertainties of the helium abundances and the metallicities make it difficult to draw definite conclusions on the true relation.

\begin{figure}
    \centering
    \includegraphics[width=\columnwidth]{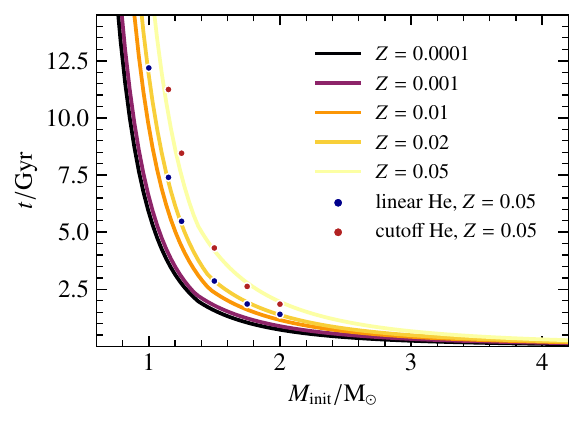}
    \caption{Lifetime function of stars until they reach the post-AGB phase as a function of their initial mass. The lines are the fit relations to the models of \citet{miller_bertolami16} with original metallicities of $Z = 0.0001$, 0.001, 0.01, and 0.02, as in \cref{fig:lifetime}. Here the extrapolation of these fits to $Z=0.05$ is also shown.
    The points are the results of newly run models following the procedure of \citet{miller_bertolami16} for $Z=0.05$, and the lines are the fit relations to those data points. The linear helium abundance is at $Y = 0.345$ and the cutoff helium abundance at $Y = 0.285$.
    The thick lines are at the same metallicities as the ones from the original models ($Z = 0.0001$, 0.001, 0.01, and 0.02) and the thin lines are at $Z = 0.00001$, 0.0003, 0.003, and 0.014, where the latter three values are the logarithmic midpoints of the original metallicities.}
    \label{fig:lifetime_helium}
\end{figure}

For these reasons, we ran two additional models following the procedure by \citet{miller_bertolami16} with a super-solar metallicity of $Z = 0.05$. The first assumes the same linear relation between the helium abundance $Y$ and metallicity $Z$ of $Y = 0.245 + 2Z$ (i.e., $Y = 0.345$, \emph{linear He}), whereas the second assumes that a saturation of helium is reached at the linearly determined value at $Z = 0.02$ (i.e., $Y = 0.285$, \emph{cutoff He}).
The lifetime functions of the MS and post-MS for stars with these initial abundances as a function of the initial stellar mass are shown in \cref{fig:lifetime_helium} as the blue (linear He) and red (cutoff He) data points. Additionally, the values from the analytic relation obtained in \cref{sec:lifetime} are shown as lines for the original four metallicities as well as for the extrapolated metallicity $Z = 0.05$.
We find that the blue linear helium data points at $Z=0.05$ lie very close to the relation at the lower metallicity $Z=0.02$ (light-orange line), whereas the red cutoff helium data points at $Z=0.05$ also correspond to the extrapolated analytical function at $Z=0.05$ (yellow line).

The newly run models show that the lifetime not only depends on metallicity, but is also heavily influenced by the helium abundance at $Z = 0.05$. Another finding is that the overall trend of the lifetimes increasing with metallicity is not valid when extrapolating the linear relation between $Y$ and $Z$ beyond $Z = 0.02$. This can be qualitatively explained by the following: in general, an increase of $Z$ leads to higher bound-free opacities in the star, which lowers the radiative energy transport and overall luminosity. More metal-rich stars thus spend more time on the MS and post-MS.
This is the case at lower metallicities, where the hydrogen abundance only minimally increases with higher $Z$.
However, at higher metallicities, a relative increase of $Z$ results in a significantly larger change of the helium abundance following the linear relation of $Y = 0.245 + 2Z$. As the hydrogen abundance is given by $X = 1 - Y - Z$, the higher value of $Y$ means that the total amount of available hydrogen is decreased, resulting in less fuel for the MS and thus shorter MS lifetimes.
Moreover, a higher helium abundance implies a higher mean molecular weight of the stellar material. Due to the high sensitivity of luminosity on the molecular weight ($L \propto \mu^4$; \citealp{kippenhahn+13}), this leads to a larger luminosity and faster fuel consumption.
In summary, when extrapolating the linear relation between $Y$ and $Z$, the stellar lifetime increases with metallicity until the rising helium abundance counteracts that effect, leading to similar lifetimes at $Z = 0.05$ and $Z = 0.02$.
In contrast, letting the helium abundance saturate and stay constant for values of $Z \geq 0.02$ means that the lifetimes continue to increase with metallicity.

For the saturated (cutoff) helium abundance, it is interesting that the simulated MS and post-MS lifetimes are overall very similar to the analytically computed values at $Z=0.05$ for stars with initial masses $M_\mathrm{init} \gtrsim \SI{1.5}{\Msun}$, but slightly higher at lower masses. This indicates that the extrapolation based on the lower metallicities (i.e., the analytic function that was fit to the original data points) corresponds to a slightly higher helium abundance than the saturated helium simulation. A thorough analysis of how the helium abundances affect the stellar lifetimes and the resulting PN populations is beyond the scope of this paper, however. It will be the subject of a future study.

Due to the poor observational constraints with respect to the $Y$-$Z$ relation, we conclude that the analytically derived lifetime function from \cref{sec:lifetime} at super-solar metallicities is still consistent with the stellar evolution simulations within the uncertainties of the observed helium abundances. Moreover, there is recent observational evidence for the helium abundance to saturate beyond a certain metallicity as measured for \object{Omega Centauri} \citep{clontz+25}.
First evidence supporting the usage of this function even at high metallicities is presented by \citet{valenzuela+24:iaus}, where the fiducial model of PICS was applied to several galaxies from a cosmological simulation, spanning a wide range of star formation histories, ages, and metallicities. The thereby obtained PN populations and their PNLF are consistent with observations and give a robust view of PNe, as will be fully shown in detail in the next paper of this series (Valenzuela et al.\ in prep.). In a future study we will further investigate how different assumptions regarding the helium abundances affect the PN populations of galaxies with different star formation histories in detail.
Comparing the observed PN populations with modeled PNe based on simulated galaxies may even provide an independent way of probing the production rate of helium at high metallicities.
In the following sections we apply the analytic function from \cref{sec:lifetime} for all metallicities.

\section{Single stellar population PNLFs}
\label{sec:ssp_pnlfs}

In \cref{sec:metallicity} we have highlighted the importance of the metallicity and helium abundance when considering some of the individual blocks required to model the PNe within a stellar population. We now showcase how the SSP age and metallicity affect the resulting PNe by exploring the parameter space for the fiducial PICS model. For this, we paired ten SSP ages between $t_\mathrm{age} = \num{0.25}$ and \SI{13}{\giga\year} with six metallicities between $Z = 0.0001$ and \num{0.08} to run PICS on and obtain corresponding PNLFs. To overcome the stochastic nature of the PN model itself with respect to the post-AGB age and the absorbing factor that determines the opacity of the nebula, we ran PICS on \num{e6} stars for each pair of age and metallicity, with equally spaced post-AGB ages between 0 and \SI{30000}{\year}.

\subsection{PNLFs with nebular metallicity correction}

\begin{figure*}
    \centering
    \includegraphics[width=0.98\textwidth]{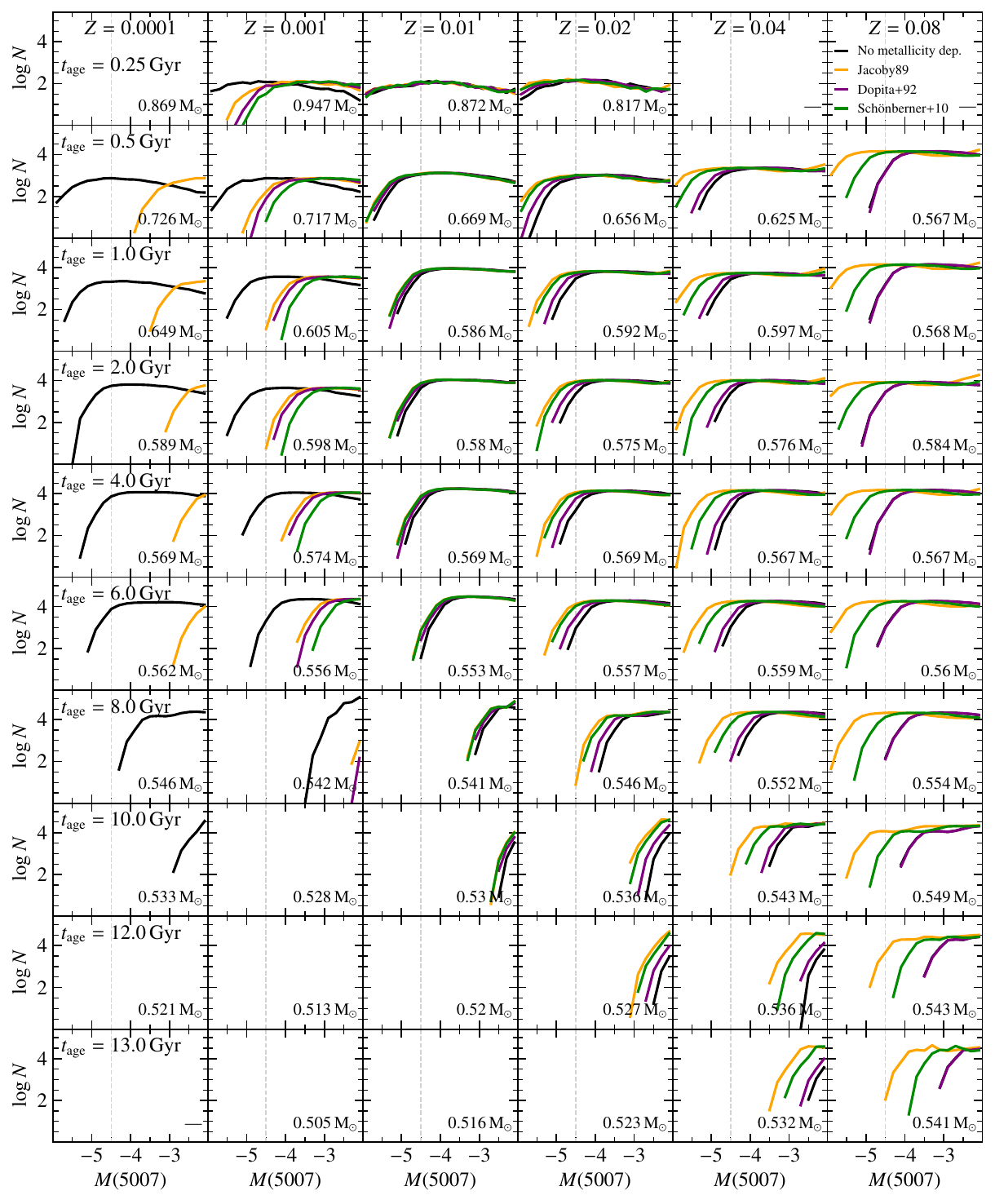}
    \caption{Single stellar population PNLFs depending on the age and metallicity for the fiducial model without circumnebular extinction, varying the nebular metallicity correction applied. Rows correspond to the different ages and columns to the metallicities. The final masses are computed from the IFMR of \citet{miller_bertolami16} and are stated in the lower right corner of each panel. In the cases where the initial mass values lie outside the modeled range, only a dash is shown (this is only the case for two panels in the top right and one in the bottom left).
    The PNLFs are obtained from \num{e6} stars of the respective final mass with equally spaced post-AGB ages between 0~and \SI{30000}{\year}. For the black PNLFs no nebular metallicity correction factor was applied, whereas the colored lines include the corrections by \citet{jacoby89:pnlfI}, \citet{dopita+92}, and \citet{schoenberner+10}.
    The bright-end cutoff is marked by the vertical gray dashed line at $M(5007) = -4.5$.
    }
    \label{fig:pnlf_grid_nebular_metallicity}
\end{figure*}

As a first step, \cref{fig:pnlf_grid_nebular_metallicity} shows the PNLFs without accounting for circumnebular extinction for a grid of SSPs with different ages (constant age per row) and metallicities (constant metallicity per column), that is the $M(5007)$ histograms for the \num{e6} PNe. The colors represent the approaches to correcting for the nebular metallicity of \citet{jacoby89:pnlfI}, \citet{dopita+92}, and \citet{schoenberner+10} that were presented in \cref{sec:pn_model}, while the black PNLFs were created from PNe without applying such a correction and functions as a reference for what the corrections are doing. Statistically, the PNLFs in a given column are always shifted by the same amount relative to each other due to the way the metallicity corrections are implemented as a function solely of metallicity. For each panel, the final mass of the central stars is stated in the bottom right corner.

The effect of accounting for nebular metallicity is immediately obvious when comparing the left to the right columns: at low metallicities, the \OIII{} brightness drops considerably (colored lines), removing any PNe at $Z=0.0001$ from bright magnitudes of $M(5007) < -4$ compared to the metallicity-independent nebular model (black line), while at high metallicities the correction leads to overall brighter PNe.
The corrections are close to zero at $Z=0.01$, which is due to the chosen base metallicity of $Z_\mathrm{base} = 0.007$ at which the corrections are defined to not change the \OIII{} brightness.
Among the three metallicity correction relations, the simple square root law by \citet{jacoby89:pnlfI} always leads to the brightest PNe (yellow). At high metallicities, the relation by \citet{dopita+92} enhances the PN brightness the least, as is expected from \cref{app:nebular_metallicity}, and again reaches a correction factor of 1 at around $Z=0.08$, such that the purple and black lines are nearly the same in the last column of \cref{fig:pnlf_grid_nebular_metallicity}.

The first finding is that not all combinations of age and metallicity contribute to the bright end of the PNLF at around $M(5007) = -4.5$ \citep[e.g.,][]{ciardullo12}, marked by the vertical gray dashed lines. Especially old stellar populations with ages of $t_\mathrm{age} \gtrsim 8$--\SI{10}{\giga\year} generally produce fainter PNe. Note, however, that when nebular metallicity effects are not taken into account, the model predicts much brighter PNe at $Z = 0.0001$ and $Z > 0.04$. The final masses of the PNe in the stellar populations not reaching the bright end for the most part lie around $M_\mathrm{final} \sim 0.51$--\SI{0.55}{\Msun}. This range lies well below the expected maximum final masses of older observed stellar population from previous studies needed to explain the brightest PNe \citep[e.g.,][]{mendez17, valenzuela+19}.

For intermediate-age stellar populations with $t_\mathrm{age} \sim 2$--\SI{8}{\giga\year}, the bright end of the PNLF is well populated with PNe. Brighter nebulae are found for lower ages. When accounting for the nebular metallicity (colored lines), this is only true for intermediate to high metallicities of $Z \gtrsim 0.001$. Even at $Z=0.001$, the older intermediate-age stellar populations ($t_\mathrm{age} \gtrsim 6$--\SI{8}{\giga\year}) do not reach the bright end. Interestingly, there is a maximum of the non-metallicity-corrected PNLF between $M(5007) = -4.5$ and $-4.0$ for a number of the age-metallicity combinations in this intermediate-age regime, with a slow decline of the PNLF towards fainter PNe. This maximum is simply shifted to other magnitudes for the metallicity corrections. All of the SSP PNLFs here feature the recognizable typical shape of observed PNLFs, just with shifts towards brighter or fainter magnitudes.
For the intermediate ages, the final masses lie between around 0.55 and \SI{0.59}{\Msun}. This approximately lies in the expected maximum central star final mass range for intermediate-to-old stellar populations as found by \citet{valenzuela+19} based on modeling PNe in mock stellar populations.

The younger stellar populations at $t_\mathrm{age} < \SI{0.5}{\giga\year}$ only feature the typical PNLF shape at the higher metallicities when not accounting for nebular metallicity (black lines). In contrast, lower-metallicities lead to much brighter PNe existing even well beyond $M(5007) = -5.0$. At high metallicities, the nebular metallicity corrections of \citet{jacoby89:pnlfI} and \citet{schoenberner+10} also lead to such bright PNe. The absolute numbers are overall much lower than for older stellar populations, which is a result of the much faster evolution along the post-AGB tracks for these more massive central stars ($M_\mathrm{final} \gtrsim 0.6$). The stars reside on the heating track for only a short period of time (see the widely spaced data points in \cref{fig:postagb_tracks_m5007} for the more massive central stars) before cooling down and dropping in luminosity, thus making bright nebulae around those massive stars rather rare.

\subsection{PNLFs with circumnebular extinction}

\begin{figure*}
    \centering
    \includegraphics[width=0.98\textwidth]{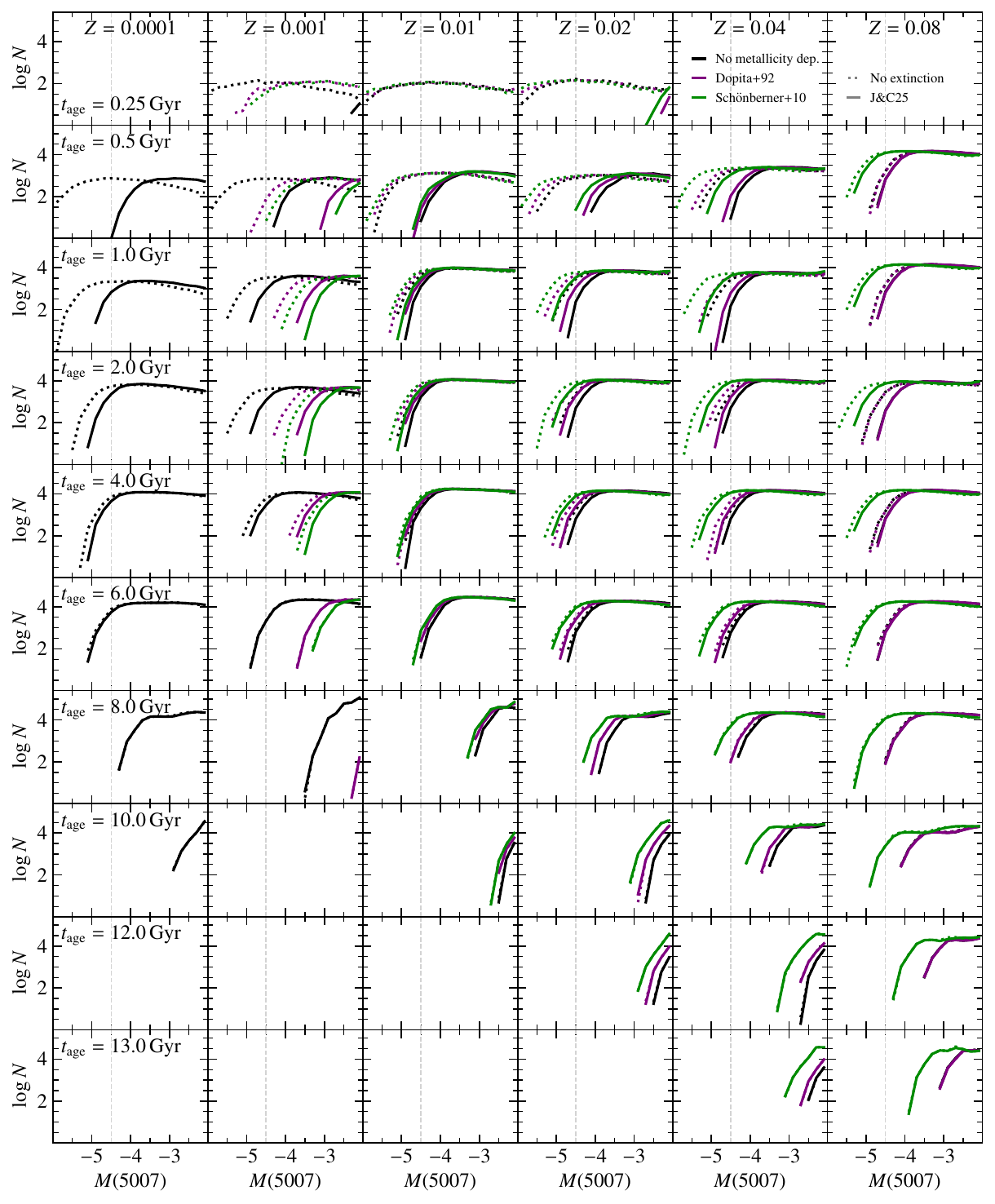}
    \caption{Single stellar population PNLFs depending on the age and metallicity for the fiducial model with different nebular metallicity corrections applied and showing the effect on the PNLFs from circumnebular dust extinction. The structure of the grid and the amount of PNe sampled for each PNLF are the same as in \cref{fig:pnlf_grid_nebular_metallicity}.
    For the black PNLFs no nebular metallicity correction factor was applied, whereas the colored lines include the corrections by \citet{dopita+92} and \citet{schoenberner+10}. The dotted PNLFs are the same as those shown in \cref{fig:pnlf_grid_nebular_metallicity} and the solid PNLFs additionally take into account the circumnebular extinction from \citet{jacoby&ciardullo25} as described in \cref{sec:circumnebular_extinction}.
    }
    \label{fig:pnlf_grid_extinction}
\end{figure*}

The fact that the PNLFs discussed in the previous section are able to reach magnitudes much brighter than the observed cut-off value is very likely due to the PN model of \citet{valenzuela+19} not incorporating circumnebular dust extinction. The circumnebular dust extinction has been observed to rise with increasing PN brightness, thus pushing the brightest PNe back down to the bright end cutoff \citep[e.g.,][]{davis+18, jacoby&ciardullo25}.
In \cref{fig:pnlf_grid_extinction} we show the full grid of PNLFs with respect to their unextincted and dust-extincted \OIII{} magnitudes. The unextincted PNLFs (dotted) are the same as those shown in \cref{fig:pnlf_grid_nebular_metallicity} for the non-metallicity dependent PNLFs (black) and those corrected for nebular metallicity (colored) through the relations of \citet{dopita+92} and \citet{schoenberner+10}. We do not show the PNLF by \citet{jacoby89:pnlfI} to keep the figure from being too cluttered and since the simple square root relation is most likely the least accurate of the three.
The solid PNLFs move to dimmer magnitudes by taking into account time-dependent dust extinction according to the Combined OR linear relation for circumnebular extinction, which is stronger in the early post-AGB phase. As a result, the extinction is seen the strongest for the fast evolving PNe of younger ages (top rows), where for $t_\mathrm{age} = \SI{0.25}{\giga\year}$ none of the PNe reach the brightest two magnitudes of the universal PNLF anymore ($M(5007) < -2.5$). At older ages of $t_\mathrm{age} \gtrsim \SI{6}{\giga\year}$, there is practically no more extinction that can be observed: the solid and dotted PNLFs are almost perfectly overlayed.

For lower-to-intermediate ages of 1--\SI{4}{\giga\year}, the extinction affecting the brightest PNe is on the order of 0.25--\SI{0.75}{\mag} and also varies slightly with metallicity. This is expected as the lifetime function and the IFMR both affect the resulting final masses and thus how fast they evolve.
\Cref{fig:pnlf_grid_extinction} nicely shows that circumstellar extinction is able to lower the SSP cutoff magnitude to the actually observed cutoff value of $M(5007) = -4.5$ for the age range of 1--\SI{4}{\giga\year}. At younger ages with generally higher final masses, the adopted prescription leads to much higher extinction for the brightest PNe and is on the order of 1.5--\SI{2.5}{\mag} for all but the highest metallicity. Only at $Z = 0.08$ the lower final masses at young-to-intermediate ages lead to significantly lower circumnebular extinction of the bright end.
The extent to which this is true naturally depends on the nebular metallicity correction applied, as for example the relation by \citet{schoenberner+10} leads to significantly brighter PNe at the metal-rich end of $Z \gtrsim 0.02$. Even the circumnebular extinction recipe does not bring down the brightest PNe to the bright end for intermediate-aged metal-rich populations ($t_\mathrm{age} \sim 0.5$--\SI{6}{\giga\year}, $Z \gtrsim 0.02$) for \citet{schoenberner+10}. We conclude that the fiducial model involving the nebular model by \citet{valenzuela+19} with the correction by \citet{dopita+92} is the most realistic with respect to our expectations from observations as it properly accounts for metallicity and simultaneously does not over-enhance the brightness of PNe at high metallicities.

As a result, we find that correcting for nebular metallicity through the relation of \citet{dopita+92} and additionally accounting for circumnebular extinction \citep{jacoby&ciardullo25} leads to a grid of PNLFs that are in general consistent with observations with respect to the bright end. They offer the basis for interpreting the observed PNLFs of composite stellar populations. At the same time, the fact that the PNLFs vary significantly between different nebular metallicity corrections also indicates that improved modeling will be necessary to decrease the systematic uncertainties seen between methods. Developing a refined nebular model is beyond the scope of this work, however.

\subsection{PNLFs with different IFMRs}

\begin{figure*}
    \centering
    \includegraphics[width=0.98\textwidth]{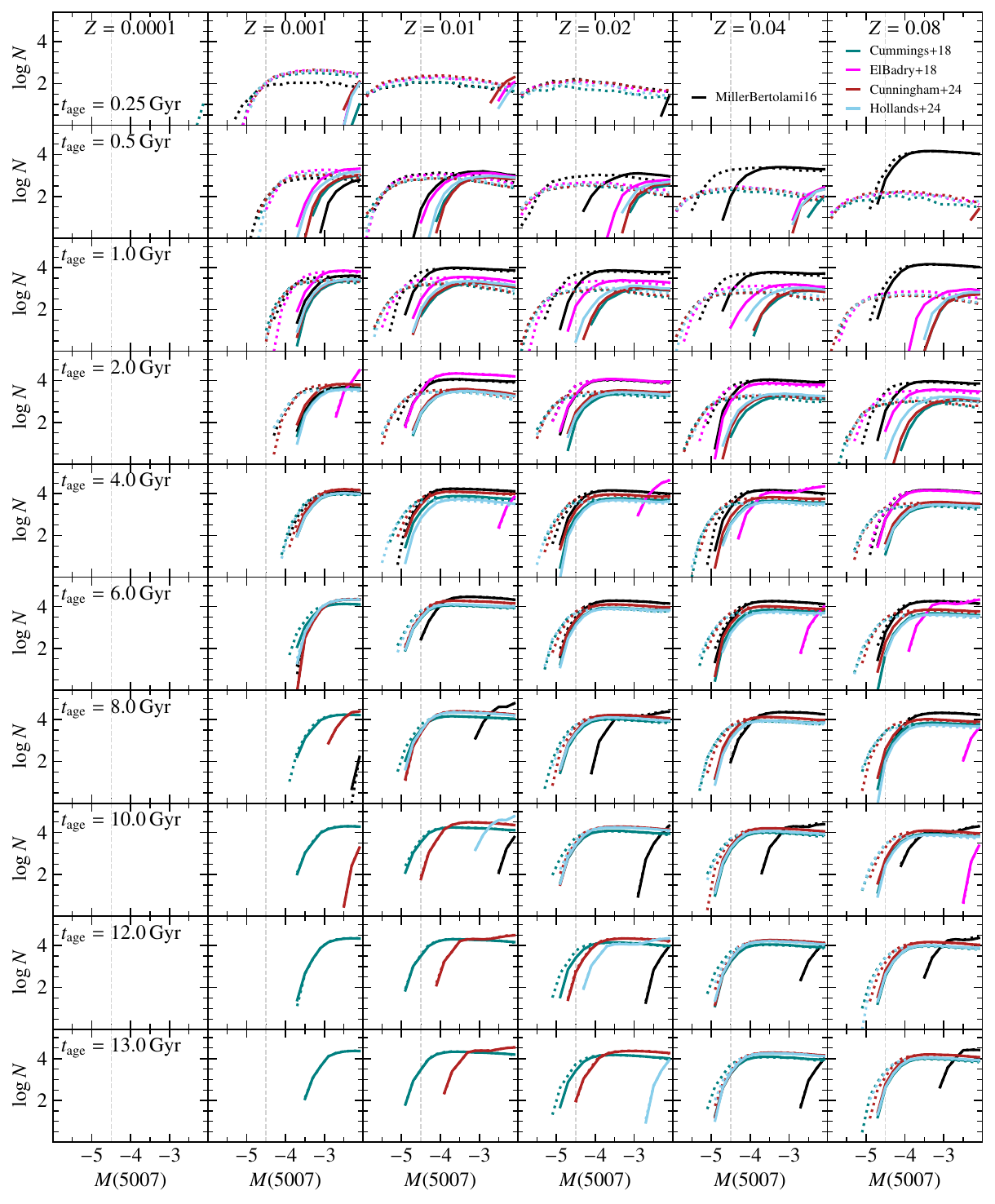}
    \caption{Single stellar population PNLFs depending on the age and metallicity for the fiducial model using the nebular metallicity correction by \citet{dopita+92} and with variations of the IFMR (colored). The IFMRs are from \citet{miller_bertolami16} in black, and \citet{cummings+18}, \citet{el_badry+18}, \citet{cunningham+24}, and \citet{hollands+24} in turquoise, magenta, red, and light blue, respectively. The structure of the grid and the amount of PNe sampled for each PNLF are the same as in \cref{fig:pnlf_grid_nebular_metallicity}. The dotted PNLFs are constructed from the intrinsic \OIII{} magnitudes and the solid PNLFs additionally take into account the circumnebular extinction from \citet{jacoby&ciardullo25} as described in \cref{sec:circumnebular_extinction}.
    }
    \label{fig:pnlf_grid_ifmr}
\end{figure*}

In \cref{sec:ifmr} we presented several IFMRs in addition to the one by \citet{miller_bertolami16} used in the fiducial PICS model, namely those by \citet{cummings+18}, \citet{el_badry+18}, \citet{cunningham+24}, and \citet{hollands+24}.
In \cref{fig:pnlf_grid_ifmr} we show the consequence of adopting an IFMR that does not depend on metallicity by using the latter four IFMRs instead of that of the fiducial PICS model by \citet{miller_bertolami16}. The black lines are the same as the blackberry lines in \cref{fig:pnlf_grid_extinction} from the fiducial PICS model with the nebular metallicity correction by \citet{dopita+92}, while the colored lines correspond to the obtained PNLFs using the IFMRs of \citet{cummings+18}, \citet{el_badry+18}, \citet{cunningham+24}, and \citet{hollands+24}.
The different IFMRs only show a similar PNLF for younger ages of around $t_\mathrm{age} \lesssim \SI{6}{\giga\year}$, and only the PICS fiducial IFMR shows a deviating behavior at higher metallicities in this age range with a larger number of PNe, but at the same time a fainter bright end of the PNLF.
In the context of circumnebular extinction, as the final masses now vary even within a single panel based on which IFMR was used, the effect of extinction is different for each PNLF because of the final mass influencing how fast the central star evolves. This can especially be seen for the young metal-rich panels, where the dotted and solid black lines lie almost on top of each other for the fiducial model using the IFMR by \citet{miller_bertolami16}, while the other IFMRs lead to much higher final masses, thus resulting in significantly stronger extinction at the bright end (the colored solid lines lie much further to the right than their dotted counterparts).

The values of age and metallicity where similar PNLFs are found for all IFMRs roughly correspond to the initial mass range of $M_\mathrm{init} \gtrsim \SI{1.3}{\Msun}$ in which the alternative IFMRs do not lead to higher final masses than at even the lowest metallicities of the fiducial models (\cref{fig:ifmr}). In contrast, for lower initial masses and thus for older stellar populations, \cref{fig:ifmr} shows that in particular the IFMR by \citet{cummings+18} surpasses the final mass values of the fiducial model at all metallicities.
At the lowest initial masses around $M_\mathrm{init} \approx \SI{0.9}{\Msun}$, both the models of \citet{cummings+18} and \citet{cunningham+24} show higher final masses than the fiducial models because their values result from linear fits in this mass range without a drop towards lower initial masses (in contrast to the fiducial model and that by \citealp{hollands+24}). Since this leads to final masses of $M_\mathrm{final} \gtrsim \SI{0.55}{\Msun}$, even such old low-mass stars are able to produce bright PNe, as seen by the bright PNe in turquoise and red in \cref{fig:pnlf_grid_ifmr} for older systems of $t_\mathrm{age} \gtrsim \SI{8}{\giga\year}$. Overall, the PNe produced with the IFMR by \citet{cummings+18} can be the brightest at older ages. The PNLFs obtained with their IFMR are surprisingly similar in bright end and shape across most ages and metallicities with only minor shifts of the bright end.
Among the other four IFMRs, differences in the bright end cutoff of up to half a magnitude can be found between the different IFMRs for $t_\mathrm{age} = \SI{6}{\giga\year}$, one magnitude at \SI{8}{\giga\year}, and more than 2--3 magnitudes in stellar populations older than \SI{10}{\giga\year}.
Finally, for the IFMR by \citet{el_badry+18} the especially low final masses obtained from initial masses of $M_\mathrm{init} \lesssim \SI{2}{\Msun}$ lead to PNe dropping out of the brightest magnitudes of the PNLF for $t_\mathrm{age} \gtrsim \SI{4}{\giga\year}$ for metallicities approximately up to the solar value (a lack of magenta PNLFs in the center to bottom left panels). At higher metallicities this is the case for $t_\mathrm{age} \gtrsim \SI{6}{\giga\year}$ ($Z = 0.04$) and $t_\mathrm{age} \gtrsim \SI{8}{\giga\year}$ ($Z = 0.08$) as a result of the longer lifetimes of metal-rich stars leading to higher initial masses at a given age.

Despite the sharp drop in the IFMR of \citet{hollands+24} towards lower masses in the range of $M_\mathrm{init} \lesssim \SI{1}{\Msun}$, which is even more extreme than the fiducial model shows, the light blue lines for \citet{hollands+24} at high ages and metallicities show that bright PNe can still be produced (bottom right panels in \cref{fig:pnlf_grid_ifmr}). This is a consequence of the alternative IFMRs not depending on metallicity, while the metallicity still influences the lifetime function, where higher metallicities lead to more massive stars reaching the post-AGB phase for a given SSP age. Because of the drastic drop of the final mass with lower initial masses, old low-metallicity SSPs do not produce bright PNe when using the IFMR by \citet{hollands+24}, leading to a lack of the light blue PNLFs in the bottom left panels in \cref{fig:pnlf_grid_ifmr}.

For the younger ages at high metallicities (top right panels), it may seem curious that all alternative IFMRs lead to similar bright PNLFs before accounting for circumnebular extinction, whereas the fiducial IFMR by \citet{miller_bertolami16} leads to a slightly dimmer bright end, but a significantly higher number of PNe. The reason for this can be understood by considering \cref{fig:ifmr}. For young ages, the initial masses are above \SI{2}{\Msun}. In this range, the high metallicity fiducial IFMRs show much lower final masses than the alternative IFMRs. The final masses are still high enough to produce relatively bright PNe, but not quite as bright as those obtained with higher final masses. However, the stars with lower final masses move more slowly across the Hertzsprung-Russell diagram before cooling down, thus allowing the nebula to survive for a longer period of time. The result is a larger number of PNe that can be observed.

The grid of PNLFs with different assumed IFMRs shows that the assumed IFMR strongly affects the PNLF that can be expected from a given stellar population, in particular for older systems. It is interesting that it is the regime of old stellar populations, where the abundance of PNe at the bright end has been a puzzle for a long time, that shows the highest sensitivity to the IFMR. This shows that understanding the IFMR dependence on metallicity (i.e., understanding the dependence of cold winds on metallicity) will play a significant role in the understanding of PNe and the PNLF in old stellar populations.
We also note that here no potential intrinsic scatter of the IFMR is accounted for, which will also affect the resulting PN populations and the brightest PNe through the variations of the final mass \citep[see also the discussion by][]{jacoby&ciardullo25}.
Finally, even though the metallicity is not considered for the alternative IFMRs themselves, the metallicity dependence of the lifetime function and the post-AGB tracks leads to a shift of the bright end of the PNLFs, which can mostly be summarized as having a brighter PNLF cutoff at higher metallicities.

\subsection{Normalized PNLFs}
\label{sec:normalized_pnlfs}

\begin{figure*}
    \centering
    \includegraphics[width=0.98\textwidth]{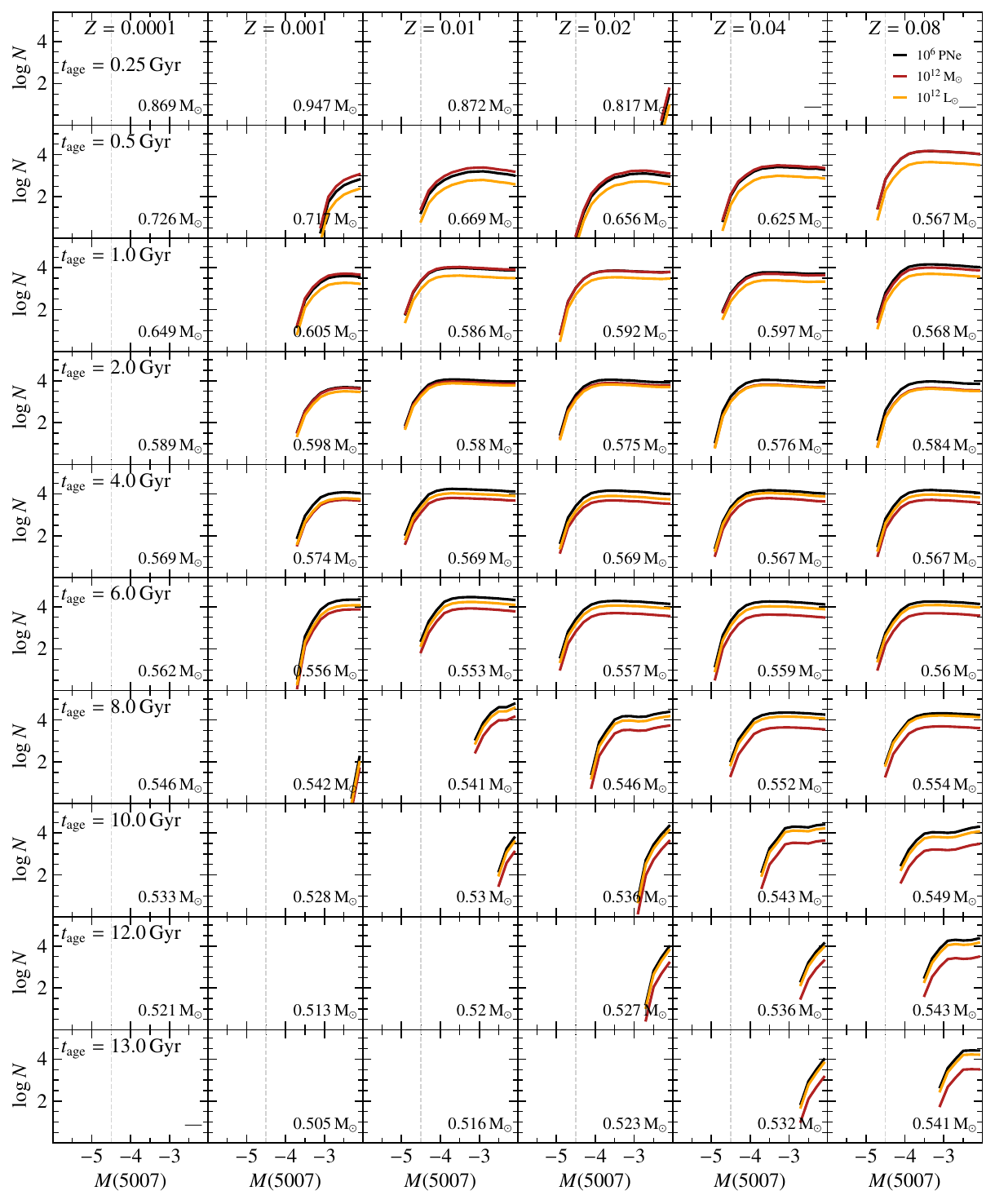}
    \caption{Single stellar population PNLFs depending on the age and metallicity for the fiducial model with the nebular metallicity correction by \citet{dopita+92} and circumnebular extinction by \citet{jacoby&ciardullo25}, normalized to different quantities. 
    The structure of the grid and the amount of PNe sampled for each PNLF are the same as in \cref{fig:pnlf_grid_nebular_metallicity}.
    The black lines are the PNLFs obtained from \num{1e6} stars of the respective final mass with equally spaced post-AGB ages between 0~and \SI{30000}{\year}.
    The red PNLFs are normalized to a total mass of \SI{e12}{\Msun} assuming a Chabrier IMF \citep{chabrier03:imf}.
    The orange PNLFs are normalized to a total bolometric luminosity of \SI{e12}{\Lsun} based on the GALAXEV CB07 SSP models by \citet{bruzual&charlot03:galaxev} using a Chabrier IMF.
    }
    \label{fig:pnlf_grid}
\end{figure*}

While the PNLF grids shown until this point were each constructed for a fixed number of \num{e6}~PNe and are useful for investigating the parameter space of the PICS model, their normalization still lacks the direct comparability to observations with respect to the underlying stellar populations.
Since the PN central stars have different initial and final masses depending on the age and metallicity (the final masses are noted in the bottom left corner of each panel in \cref{fig:pnlf_grid_nebular_metallicity}), varying numbers of PNe are expected to be found if the total SSP mass or luminosity is kept constant, based on the IMF.
In \cref{fig:pnlf_grid} we show the \citet{dopita+92} nebular metallicity-corrected and circumnebular-extincted PNLFs for a fixed number of \num{e6}~PNe as shown in the other PNLF grids (black), for a fixed total stellar mass of $M_* = \SI{e12}{\Msun}$ assuming a \citet{chabrier03:imf} IMF (red), and for a fixed total bolometric luminosity of $L_\mathrm{bol} = \SI{e12}{\Lsun}$ based on the GALAXEV CB07 SSP models by \citet{bruzual&charlot03:galaxev} using a \citet{chabrier03:imf} IMF (orange).

For the luminosity-normalized PNLFs (orange), there are relatively more PNe compared to the fixed-number PNLFs (black) for older ages. The reason for this arises from the fact that younger stellar populations are intrinsically brighter than older ones, thus resulting in less total mass when keeping the total bolometric luminosity constant. In turn, this leads to a smaller number of PNe for young ages.
The normalization to bolometric luminosity also gives us the possibility to determine the luminosity-specific PN numbers, typically denoted by $\alpha$. As our model produces complete samples of PNe for a given SSP, it is possible to compute the actual value of $\alpha$ without having to assume an underlying PNLF shape based on which the number is estimated by extrapolation. Instead, the $\alpha$ value is obtained directly from counting the number of PNe in a given SSP brighter than a given magnitude. As the model can produce extremely dim magnitudes for objects that are essentially not PNe anymore, we limited the $\alpha$ determination to PNe with $M(5007) \leq +3.5$, which corresponds to the 8 brightest magnitudes with an assumed bright-end cutoff of $M(5007) = -4.5$. We chose this magnitude range to align with the estimation of \citet{buzzoni+06} on PNe spanning roughly 8 orders of magnitude.
The values of $\alpha$ for the same grid of SSPs found in \cref{fig:pnlf_grid} are shown in \cref{tab:alpha}, using the extincted PICS fiducial model and the same bolometric luminosity prescription by \citet{bruzual&charlot03:galaxev} as described above.
Note that this approach differs from that of \citet{buzzoni+06} through our full modeling of PNe: we can simply count the PNe and do not have to make assumptions on the specific evolutionary flux, but rather can infer it from the model results themselves (see below).

\begin{table}
    \centering
    \caption{Luminosity-specific PN numbers, $\alpha$ in units of \SI{e-7}{\per\Lsun}, of the SSP PNLFs for the brightest 8 magnitudes below the typical bright-end cutoff of $M(5007) = -4.5$. Each row corresponds to the given age $t_\mathrm{age}$ and each column to the given metallicity $Z$. The values were computed from the constant luminosity PNLFs shown in \cref{fig:pnlf_grid} using the extincted PICS fiducial model with metallicity correction by \citet{dopita+92}, circumnebular extinction by \citet{jacoby&ciardullo25}, and the same bolometric luminosity prescription by \citet{bruzual&charlot03:galaxev}.}
    \label{tab:alpha}
    \begin{tabular}{cS[table-format=1.3,table-auto-round=true]S[table-format=1.3,table-auto-round=true]S[table-format=1.3,table-auto-round=true]S[table-format=1.3,table-auto-round=true]S[table-format=1.3,table-auto-round=true]S[table-format=1.3,table-auto-round=true]}
      \hline
      $t_\mathrm{age}$ & \multicolumn{6}{c}{$Z$} \\
      (\si{\giga\year}) & {0.0001} & {0.001} & {0.01} & {0.02} & {0.04} & {0.08} \\
      \hline\\[-0.2cm]
      0.25 & {--} & 0.42854 & 2.03556 & 2.07525 & {--} & {--} \\
      0.5 & 0.0331 & 1.41918 & 2.84498 & 3.34799 & 3.12508 & 2.66922 \\
      1.0 & 0.13838 & 2.46699 & 3.55512 & 4.2242 & 3.60161 & 3.15745 \\
      2.0 & 0.43228 & 3.58044 & 5.51148 & 5.38958 & 5.0925 & 3.92517 \\
      4.0 & 0.82489 & 3.34094 & 5.41347 & 5.10111 & 6.6418 & 5.45984 \\
      6.0 & 1.37045 & 3.6553 & 5.24527 & 5.48558 & 5.32632 & 6.08398 \\
      8.0 & 2.15083 & 4.98603 & 5.76273 & 5.64002 & 5.84433 & 7.02519 \\
      10.0 & 1.82507 & 3.88037 & 5.75783 & 5.89184 & 5.93659 & 5.34369 \\
      12.0 & {--} & {--} & {--} & 6.39059 & 6.14773 & 5.64185 \\
      13.0 & {--} & {--} & {--} & {--} & 6.27679 & 5.78672 \\
      \hline\\[-0.3cm]
      & \multicolumn{6}{c}{$\alpha$ in units of \SI{e-7}{\per\Lsun}} \\
      \hline
    \end{tabular}
\end{table}

\begin{table}
    \centering
    \caption{PN visibility lifetimes, $\tau_\mathrm{PN}$ in units of \SI{e3}{\year}, of the PNe resulting from the SSPs for the brightest 8 magnitudes below the typical bright-end cutoff of $M(5007) = -4.5$. Each row corresponds to the given age $t_\mathrm{age}$ and each column to the given metallicity $Z$.}
    \label{tab:tau_pn}
    \begin{tabular}{cS[table-format=2.2,table-auto-round=true]S[table-format=2.2,table-auto-round=true]S[table-format=2.2,table-auto-round=true]S[table-format=2.2,table-auto-round=true]S[table-format=2.2,table-auto-round=true]S[table-format=2.2,table-auto-round=true]}
      \hline
      $t_\mathrm{age}$ & \multicolumn{6}{c}{$Z$} \\
      (\si{\giga\year}) & {0.0001} & {0.001} & {0.01} & {0.02} & {0.04} & {0.08} \\
      \hline\\[-0.2cm]
      0.25 & {--} & 5.19552 & 21.5351 & 20.6043 & {--} & {--} \\
      0.5 & 0.26652 & 11.8565 & 22.0854 & 23.8652 & 24.7297 & 26.8588 \\
      1.0 & 0.86355 & 15.8165 & 24.9008 & 26.3888 & 25.9456 & 26.8361 \\
      2.0 & 2.53794 & 16.8331 & 25.2902 & 27.0936 & 27.0185 & 26.2649 \\
      4.0 & 5.32125 & 19.3369 & 26.3403 & 27.2713 & 27.2244 & 26.859 \\
      6.0 & 7.7826 & 20.4164 & 26.9032 & 27.2912 & 27.2422 & 26.8781 \\
      8.0 & 11.6942 & 26.5375 & 27.2346 & 27.3686 & 27.1521 & 26.8479 \\
      10.0 & 9.52782 & 19.428 & 26.6786 & 27.6364 & 27.4307 & 26.8175 \\
      12.0 & {--} & {--} & {--} & 28.3178 & 27.5861 & 27.1392 \\
      13.0 & {--} & {--} & {--} & {--} & 27.547 & 27.2826 \\
      \hline\\[-0.3cm]
      & \multicolumn{6}{c}{$\tau_\mathrm{PN}$ in units of \SI{e3}{\year}} \\
      \hline
    \end{tabular}
\end{table}

These values for the most part lie between \SI{5e-7}{\per\Lsun} for solar metallicities of $Z = 0.01$--0.02 and ages of $t_\mathrm{age} \gtrsim \SI{2}{\giga\year}$, which drop towards lower metallicities and ages, and rise towards higher metallicities and ages.
These results are generally on the same order of magnitude as what \citet{buzzoni+06} found, except that they find $\alpha$ to correlate much more strongly with age and much less with metallicity. The trend with metallicity that we find is largely due to the metallicity-dependent stellar evolution models employed from \citet{miller_bertolami16} with respect to the stellar lifetimes and the IFMR, which was not taken into account by \citet{buzzoni+06}.
Our smaller trend with age, however, could arise from the assumption of the nebular model from \citet{valenzuela+19} that PNe have a maximum lifetime of \SI{30000}{\year}. While this assumption is likely sufficient for most observed PNe, the dimmest PNe may very well also occur after this time frame. This could increase the total number of PNe for older stellar populations since their central stars have lower masses and thus evolve more slowly along the stellar tracks, leading to a slower loss of temperature and luminosity.

As \citet{buzzoni+06} discuss, the $\alpha$ parameter can be linked to the PN visibility lifetime $\tau_\mathrm{PN}$. For the PICS model, it is possible to directly extract this lifetime for each age and metallicity as the time that a PN has a magnitude of $M(5007) \leq +3.5$. From the \num{e6}~modeled PNe uniformly sampled across \SI{30000}{\year} for each SSP, we took the fraction of PNe in that magnitude range as the factor to be multiplied with \SI{30000}{\year} to obtain the respective PN visibility lifetime, which we show in \cref{tab:tau_pn} for the same SSP grid.
Around solar metallicity of $Z = 0.01$--0.02, the visibility lifetimes roughly range between \num{25000} and \SI{28000}{\year} for SSP ages $t_\mathrm{age} \geq \SI{1}{\giga\year}$. Lower metallicities and ages lead to shorter PN visibility lifetimes, similar to the trends seen with $\alpha$ as expected.
Note that the values do not reach the full \SI{30000}{\year} because of the stochastic nature of the model by \citet{valenzuela+19} through the absorption parameter, which accounts for possible optical transparency of the nebula. This can drop the \OIII{} flux below the 8 magnitudes for the faintest PNe. Further, these values of the PN visibility lifetime are affected by the model's choice of \SI{30000}{\year}, a matter that will be addressed through improved nebular models in the future.
From $\alpha$ and $\tau_\mathrm{PN}$ the specific evolutionary flux can be determined as their fraction $\mathcal{B} = \alpha / \tau_\mathrm{PN}$ \citep{buzzoni+06}.
For instance, at $Z = 0.01$ and $t_\mathrm{age} = \SI{6.0}{\giga\year}$, the specific evolutionary flux is $\mathcal{B} = \SI{1.95e-11}{\per\Lsun\per\year}$, which exactly lies in the upper range shown in fig.~1 from \citet{buzzoni+06}.

In the literature the use of $\alpha_{2.5}$ has commonly been used due to the fact that observationally one is generally restricted to only the brightest magnitudes of the PNLF \citep[e.g.,][]{jacoby+90:pnlfV, longobardi+13, hartke+17}. Hereby only the 2.5 brightest magnitudes are taken into account for the luminosity-specific PN number. We show the predicted values of $\alpha_{2.5}$ and their corresponding visibility lifetimes $\tau_{\mathrm{PN},2.5}$ in \cref{app:alpha25}.

For some nearby galaxies it has been possible to measure a statistically complete PNLF down to six magnitudes below the bright cutoff \citep[e.g.,][]{reid&parker10:III, bhattacharya+19:I}, which has revealed significant variations of the PNLF shape at its faint end. We show the same grid of normalized PNLFs as in \cref{fig:pnlf_grid} up to a magnitude of $M(5007) = +2$ in \appref{app:pnlf_extended}, which is around 6.5 magnitudes below the bright end of the PNLF. These will be relevant in upcoming studies to compare the models with such nearby deep observations and provide a means for interpreting the variations of the PNLF shape.
We highlight two insights from the extended magnitude PNLF grid: first, we find the dip or \enquote{camel} shape previously found in some galaxies with recent star formation like the Magellanic Clouds or \ngc{300} \citep{jacoby&de_marco02, reid&parker10:III, pena+12, soemitro+23:IV} and simulations of mock stellar populations with more massive central stars \citep{mendez+08, valenzuela+19}. Here we find that this dip already arises from the PNLF of a sole SSP and is not necessarily the result of multiple stellar populations overlapping with each other, as has for example been proposed by \citet{rodriguez_gonzalez+15}. As proposed by \citet{ciardullo10}, the physical reason for the dip is that more massive post-AGB stars can host brighter PNe on their heating track than lower-mass stars, but their extremely fast evolution leads to them cooling down very quickly, resulting in a rapid decline of the \OIII{} emission before reaching a moderate brightness, where their evolution slows down and PNe accumulate, which is where the PNLF rises again.
Second, the differences in shape and slope found at the faint end of the PNLF in observations indicate that these individual SSP PNLFs will help better understand and interpret the different behaviors found at the faint end of the PNLF, for example for PN populations in \m{31} and its outskirts \citep{bhattacharya+21:III}.

As a final comment, since any given stellar population is made up of many SSPs, any observed or simulated PNLF is in fact a superposition of these SSP PNLFs.
In an upcoming study, we will use this fact to present a test using resolved star formation histories of observed galaxies to compare the PICS-predicted PNLF based on the observationally inferred stellar populations to the actually observed PNLF. With those tests we will demonstrate how successfully PICS is capable of modeling accurate PN populations self-consistently.

\section{Summary and conclusion}
\label{sec:conclusion}

This first paper in a series introduces PICS (PNe In Cosmological Simulations), a modular framework that models the PN population contained within an SSP. The SSP is described by its total mass, age, IMF, and optionally by its metallicity and/or further abundances. For our fiducial model, we consider the SSP metallicity mass ratio and use the metallicity-dependent lifetime function, IFMR, and post-AGB tracks from \citet{miller_bertolami16} to model the central stars of the PNe. The nebula and its \OIII{} emission are modeled using the method from \citet{valenzuela+19} and we account for the nebular metallicity according to the results by \citet{dopita+92}. Finally, time-dependent circumnebular dust extinction is applied to the \OIII{} emission derived from the observed relation by \citet{jacoby&ciardullo25}. PICS thus produces the main observable quantity of extragalactic PNe and can be employed to compare theoretical models and hydrodynamical cosmological simulations with observations, which in turn can be used to interpret the properties of the PNe and underlying stellar populations.

We show how the initial metallicity of the star affects the stellar evolution relevant to PNe: higher metallicities lead to more massive stars reaching the post-AGB phase later than their lower-metallicity counterparts, while the final masses after the AGB phase are lower at constant initial mass. Higher metallicities also lead to lower luminosities and temperatures of the central stars for a fixed final mass.
Considering both the lifetime function and the IFMR, the effect of the metallicity is strongest at ages below 3--\SI{4}{\giga\year}, where lower metallicities lead to more massive PN central stars, and at higher ages above \SI{6}{\giga\year}, where higher metallicities lead to more massive central stars (\cref{fig:metallicity_effect}). The effect on final mass becomes strongest at very young and at very old ages.
The final mass is important for the maximal brightness of $M(5007)$ that can be reached throughout the post-AGB phase and reaching higher final masses will increase the probability of finding such a PN at the bright end of the PNLF.
Furthermore, these relations provide the first indications that the metallicity is relevant for older stellar populations, in particular as these are oftentimes more metal-rich in massive elliptical galaxies than in disk galaxies like our own Milky Way.

One important finding of this work is that for high-metallicity stars, the metallicity is not the only property affecting the lifetime of stars. While higher metallicities generally do increase the opacity of the star, leading to longer lifetimes, the accompanying increase of the helium abundance counteracts this trend due to higher main sequence luminosities and less available fuel. As the relation of helium abundance with metallicity is still poorly constrained from observations, in particular at high metallicities, we explored the parameter space ranging from a linearly increasing helium abundance with metallicity to a constant helium abundance of $Y = 0.285$ above a metallicity of $Z = 0.02$. Such a constant helium abundance above a certain metallicity would be in line with recent observations of the metallicity of the stellar population of the globular cluster \object{Omega Centauri} \citep{clontz+25}.

Assuming the linear relation, the lifetime function does not lead to longer timescales at higher metallicities above $Z = 0.02$ anymore, whereas assuming a saturation of the helium abundance results in longer lifetimes at $Z > 0.02$.
The helium abundance relation is therefore especially relevant for metal-rich stellar populations, as it governs how the lifetime function behaves for the individual stars. A higher helium abundance causes shorter lifetimes, leading to less massive stars becoming PNe after a given time at fixed metallicity. This in turn means that the brightest possible magnitudes $M(5007)$ are dimmer.
If the helium abundance is capped, however, the stellar lifetimes increase, leading to more massive central stars at older ages for super-solar metallicity stellar populations.
Based on the metallicities originally used by \citet{miller_bertolami16} of $Z = 0.0001$--0.02, we derived an analytic expression for the lifetime function, whose extrapolation lies between the simulated values for linear and constant helium abundance relations, but closer to the latter. Thus, for the fiducial model, we pick the constant helium abundance relation due to it lying closer to the analytic function.
Since longer lifetimes of metal-rich stars increase the occurrence of bright-end PNe in old stellar populations, our findings indicate that a saturation of the helium abundance at high metallicities could be part of the solution for the universal bright-end cutoff of the PNLF.
The implications would be that high-metallicity stars have longer lifetimes than expected from models assuming non-saturated helium abundances. In future studies we will extend our stellar grids to include the effect of the helium abundance consistently with the prediction from cosmological models.

To obtain a complete overview of how the SSP metallicity and age affect the resulting PN population, we produced a grid of PNLFs, each corresponding to a fixed pair of metallicity and age. The PNLF shapes are generally consistent with each other and have a varying bright end cutoff: older SSPs generally contain less luminous PNe. However, at older ages above \SI{6}{\giga\year}, higher metallicities again lead to brighter PNe forming. In addition, at almost all ages the lowest metallicities of $Z = 0.0001$ and $Z = 0.001$ lead to a significant dimming of the PNe through any of the three nebular metallicity corrections we have implemented in this work \citep{jacoby89:pnlfI, dopita+92, schoenberner+10}, especially compared to not accounting for the nebular metallicity.
In particular, the brighter old and metal-rich PNe could in part explain how old elliptical galaxies still retain a similar bright end of the PNLF compared to disk galaxies as they tend to be especially metal-rich.
The grids of SSP PNLFs presented in this work set the theoretical foundation for interpreting the relations of the observed PNLF bright end with properties of the underlying stellar populations with respect to their age and metallicity.

We also present a grid of extended PNLFs down to six magnitudes below the bright end cutoff, which are needed for comparison and interpretation of PN studies of Local Universe galaxies, where it has been possible to obtain statistically complete samples even of very faint PNe, for example in the LMC or \m{31} \citep{reid&parker10:III, bhattacharya+21:III}. The extended PNLFs reveal that the \enquote{dip} or \enquote{camel shape} of the PNLF with a minimum at around $M(5007) \approx -2$--0, which has previously been seen in observations \citep[e.g.,][]{jacoby&de_marco02, reid&parker10:III, pena+12, soemitro+23:IV} and models \citep[e.g.,][]{mendez+08, valenzuela+19}, is in fact an intrinsic property of PNe with more massive central stars and arises even in SSPs. It is thereby not only a result of two or more modes of PNLFs from multiple stellar populations, as has been suggested by \citet{rodriguez_gonzalez+15}.
Of course, any real PNLF is a superposition of SSP PNLFs, which can lead to an even larger variety of observed faint PNLF shapes. This carries the potential of being another independent diagnostic tool for investigating the underlying star formation history and metallicity properties of galaxies and their formation histories, which we will also be presenting with a self-consistent test on observations in an upcoming study.

Through the PNLF grids we investigate the effects of applying different nebular metallicity corrections, which are critical to avoid overly luminous extremely metal-poor PNe. The square root law by \citet{jacoby89:pnlfI} is the relation first found for solar-like metallicities and enhances the brightness of metal-rich PNe very strongly when extrapolated, and leads to the smallest dimming of metal-poor PNe. In contrast, the relations by \citet{dopita+92} and \citet{schoenberner+10} enhance it less significantly at the metal-rich end, and the relation by \citealp{dopita+92} the least. Applying circumnebular extinction further decreases the brightness of the brightest PNe of very young stellar populations, such that the bright end of the PNLF can only be found at solar-like and higher metallicities for stellar ages of ${\gtrsim}\SI{1}{\giga\year}$.

Finally, we studied how different IFMRs affect the resulting PN populations by considering metallicity-independent IFMRs determined from the solar neighborhood in the Milky Way in addition to the relation used for the fiducial model from \citet{miller_bertolami16}. Most of the various IFMRs (except that by \citealp{el_badry+18}) do not change the SSP PNLFs much for younger stellar populations of ${\lesssim}\SI{6}{\giga\year}$. However, at larger ages the metallicity-independent IFMRs mostly lead to brighter PNe being produced, in part even reaching the bright-end cutoff for the IFMR by \citet{cummings+18}. The reason for this is that some of these IFMRs do not see the final stellar masses drop as steeply towards low initial masses as the relations by \citet{miller_bertolami16}.
Only the IFMR by \citet{el_badry+18} predicts much lower final masses in the low-mass regime, that is for old stellar ages. As a result, no PNe are produced in the brightest 4 magnitudes for ages of ${\gtrsim}\SI{6}{\giga\year}$ for this IFMR.
Furthermore, the lack of metallicity dependence means that at high metallicities the PN central stars are more massive than when employing the metallicity-dependent IFMR, in turn leading to on average brighter PNe.
The implications for the existence of bright PNe in old stellar populations are very important: The IFMR, in particular at initial masses of $M_\mathrm{init} \lesssim \SI{1.2}{\Msun}$, plays a significant role in how massive the central stars are and therefore how bright the PNe can get. Getting better constraints and developing a more robust theoretical understanding of the mass loss in the AGB phase is therefore essential for explaining the observed bright end of the PNLF in old stellar populations.

With PICS, we have laid the groundwork for populating cosmological simulations with PNe, thereby producing PN populations based on self-consistent star formation histories within a cosmological context.
On the theoretical side, this allows us to make predictions for the PNLF behavior based on statistically significant samples of galaxies with diverse morphologies and star formation histories, which will aid our understanding of the universal bright end of the PNLF.
Furthermore, the modular structure of PICS makes parameter studies possible, where controlled changes to one specific block in the pipeline can be used to understand its effect on the resulting PN population, just as done for the IFMR in this work.
For the observational side, it will be possible to compare the modeled PN populations to observations by identifying appropriate simulated galaxy analogues, and to offer an interpretation for the observed PNe and their properties.

To conclude, we have shown in this study that it is essential to account for both the metallicity and helium abundances in PNe models, as well as improving the observations and models of the IFMR. These factors all particularly affect the brightest PNe, leading us one step further in our understanding of the universal bright end of the PNLF, which we will explore in more detail in future studies.

\begin{acknowledgements}
We thank Souradeep Bhattacharya, Tadziu Hoffmann, George Jacoby, Rolf-Peter Kudritzki, and Joachim Puls for helpful comments and discussions, and Tim Cunningham for providing us with his data tables of the IFMR.

We thank the referee, Robin Ciardullo, for his constructive feedback that helped us significantly improve the paper.

LMV acknowledges support by the German Academic Scholarship Foundation (Studienstiftung des deutschen Volkes) and the Marianne-Plehn-Program of the Elite Network of Bavaria. 

M3B is partially funded by CONICET and Agencia I+D+i through grants PIP-2971 and PICT 2020-03316, and by  CONICET-DAAD 2022 bilateral cooperation grant number 80726.

This research was supported by the Excellence Cluster ORIGINS, funded by the Deutsche Forschungsgemeinschaft under Germany's Excellence Strategy -- EXC-2094-390783311.

The following software was used for this work: Julia \citep{bezanson+17:julia}, DataInterpolations.jl \citep{bhagavan+24:datainterpolations.jl}, Interpolations.jl \citep{holy+:interpolations.jl}, \code{matplotlib} \citep{hunter07:matplotlib}, and \code{numpy} \citep{harris+20:numpy}.
\end{acknowledgements}

\bibliographystyle{style/aa}
\bibliography{bib}

\begin{thebibliography}{87}
\expandafter\ifx\csname natexlab\endcsname\relax\def\natexlab#1{#1}\fi

\bibitem[{Althaus {et~al.}(2009)Althaus, Panei, Romero, Rohrmann, Córsico, {García-Berro}, \& Miller~Bertolami}]{althaus+09}
Althaus, L.~G., Panei, J.~A., Romero, A.~D., {et~al.} 2009, \aap, 502, 207

\bibitem[{Aniyan {et~al.}(2018)Aniyan, Freeman, Arnaboldi, Gerhard, Coccato, Fabricius, Kuijken, Merrifield, \& Ponomareva}]{aniyan+18}
Aniyan, S., Freeman, K.~C., Arnaboldi, M., {et~al.} 2018, \mnras, 476, 1909

\bibitem[{Aniyan {et~al.}(2021)Aniyan, Ponomareva, Freeman, Arnaboldi, Gerhard, Coccato, Kuijken, \& Merrifield}]{aniyan+21}
Aniyan, S., Ponomareva, A.~A., Freeman, K.~C., {et~al.} 2021, \mnras, 500, 3579

\bibitem[{Asplund {et~al.}(2021)Asplund, Amarsi, \& Grevesse}]{asplund+21}
Asplund, M., Amarsi, A.~M., \& Grevesse, N. 2021, \aap, 653, A141

\bibitem[{Asplund {et~al.}(2009)Asplund, Grevesse, Sauval, \& Scott}]{asplund+09}
Asplund, M., Grevesse, N., Sauval, A.~J., \& Scott, P. 2009, \araa, 47, 481

\bibitem[{Bezanson {et~al.}(2017)Bezanson, Edelman, Karpinski, \& Shah}]{bezanson+17:julia}
Bezanson, J., Edelman, A., Karpinski, S., \& Shah, V.~B. 2017, SIAM Review, 59, 65

\bibitem[{Bhagavan {et~al.}(2024)Bhagavan, de~Koning, Maddhashiya, \& Rackauckas}]{bhagavan+24:datainterpolations.jl}
Bhagavan, S., de~Koning, B., Maddhashiya, S., \& Rackauckas, C. 2024, Journal of Open Source Software, 9, 6917

\bibitem[{Bhattacharya {et~al.}(2021)Bhattacharya, Arnaboldi, Gerhard, McConnachie, Caldwell, Hartke, \& Freeman}]{bhattacharya+21:III}
Bhattacharya, S., Arnaboldi, M., Gerhard, O., {et~al.} 2021, \aap, 647, A130

\bibitem[{Bhattacharya {et~al.}(2019)Bhattacharya, Arnaboldi, Hartke, Gerhard, Comte, McConnachie, \& Caldwell}]{bhattacharya+19:I}
Bhattacharya, S., Arnaboldi, M., Hartke, J., {et~al.} 2019, \aap, 624, A132

\bibitem[{Bl{\"o}cker(1995)}]{bloecker95:tracks}
Bl{\"o}cker, T. 1995, \aap, 299, 755

\bibitem[{Bressan {et~al.}(2012)Bressan, Marigo, Girardi, Salasnich, Dal~Cero, Rubele, \& Nanni}]{bressan+12:parsec}
Bressan, A., Marigo, P., Girardi, {\relax Léo}., {et~al.} 2012, \mnras, 427, 127

\bibitem[{Bruzual \& Charlot(2003)}]{bruzual&charlot03:galaxev}
Bruzual, G. \& Charlot, S. 2003, Monthly Notices of the Royal Astronomical Society, 344, 1000

\bibitem[{Buzzoni {et~al.}(2006)Buzzoni, Arnaboldi, \& Corradi}]{buzzoni+06}
Buzzoni, A., Arnaboldi, M., \& Corradi, R. L.~M. 2006, \mnras, 368, 877

\bibitem[{Caffau {et~al.}(2011)Caffau, Ludwig, Steffen, Freytag, \& Bonifacio}]{caffau+11}
Caffau, E., Ludwig, H.~G., Steffen, M., Freytag, B., \& Bonifacio, P. 2011, \solphys, 268, 255

\bibitem[{Cardelli {et~al.}(1989)Cardelli, Clayton, \& Mathis}]{cardelli+89}
Cardelli, J.~A., Clayton, G.~C., \& Mathis, J.~S. 1989, \apj, 345, 245

\bibitem[{Casagrande {et~al.}(2007)Casagrande, Flynn, Portinari, Girardi, \& Jimenez}]{casagrande+07}
Casagrande, L., Flynn, C., Portinari, L., Girardi, L., \& Jimenez, R. 2007, \mnras, 382, 1516

\bibitem[{Chabrier(2003)}]{chabrier03:imf}
Chabrier, G. 2003, \pasp, 115, 763

\bibitem[{Ciardullo(2010)}]{ciardullo10}
Ciardullo, R. 2010, \pasa, 27, 149

\bibitem[{Ciardullo(2012)}]{ciardullo12}
Ciardullo, R. 2012, \apss, 341, 151

\bibitem[{Ciardullo {et~al.}(2002)Ciardullo, Feldmeier, Jacoby, {Kuzio de Naray}, Laychak, \& Durrell}]{ciardullo+02:pnlfXII}
Ciardullo, R., Feldmeier, J.~J., Jacoby, G.~H., {et~al.} 2002, \apj, 577, 31

\bibitem[{Ciardullo \& Jacoby(1992)}]{ciardullo&jacoby92:pnlfVIII}
Ciardullo, R. \& Jacoby, G.~H. 1992, \apj, 388, 268

\bibitem[{Ciardullo \& Jacoby(1999)}]{ciardullo&jacoby99}
Ciardullo, R. \& Jacoby, G.~H. 1999, \apj, 515, 191

\bibitem[{Ciardullo {et~al.}(1989)Ciardullo, Jacoby, Ford, \& Neill}]{ciardullo+89:pnlfII}
Ciardullo, R., Jacoby, G.~H., Ford, H.~C., \& Neill, J.~D. 1989, \apj, 339, 53

\bibitem[{Ciardullo {et~al.}(2005)Ciardullo, Sigurdsson, Feldmeier, \& Jacoby}]{ciardullo+05}
Ciardullo, R., Sigurdsson, S., Feldmeier, J.~J., \& Jacoby, G.~H. 2005, \apj, 629, 499

\bibitem[{Clontz {et~al.}(2025)Clontz, Seth, Wang, Souza, H{\"a}berle, Nitschai, Neumayer, Latour, Milone, {Feldmeier-Krause}, Kacharov, Libralato, Bellini, {van de Ven}, \& {Alfaro-Cuello}}]{clontz+25}
Clontz, C., Seth, A.~C., Wang, Z., {et~al.} 2025, \apj, 984, 162

\bibitem[{Cummings {et~al.}(2018)Cummings, Kalirai, Tremblay, {Ramirez-Ruiz}, \& Choi}]{cummings+18}
Cummings, J.~D., Kalirai, J.~S., Tremblay, P.~E., {Ramirez-Ruiz}, E., \& Choi, J. 2018, \apj, 866, 21

\bibitem[{Cunningham {et~al.}(2024)Cunningham, Tremblay, \& W.~O'Brien}]{cunningham+24}
Cunningham, T., Tremblay, P.-E., \& W.~O'Brien, M. 2024, \mnras, 527, 3602

\bibitem[{Davis {et~al.}(2018)Davis, Ciardullo, Jacoby, Feldmeier, \& Indahl}]{davis+18}
Davis, B.~D., Ciardullo, R., Jacoby, G.~H., Feldmeier, {\relax John}.~J., \& Indahl, B.~L. 2018, \apj, 863, 189

\bibitem[{Dopita {et~al.}(1992)Dopita, Jacoby, \& Vassiliadis}]{dopita+92}
Dopita, M.~A., Jacoby, G.~H., \& Vassiliadis, E. 1992, \apj, 389, 27

\bibitem[{Dopita \& Meatheringham(1991)}]{dopita&meatheringham91:II}
Dopita, M.~A. \& Meatheringham, S.~J. 1991, \apj, 377, 480

\bibitem[{Douglas {et~al.}(2002)Douglas, Arnaboldi, Freeman, Kuijken, Merrifield, Romanowsky, Taylor, Capaccioli, Axelrod, Gilmozzi, Hart, Bloxham, \& Jones}]{douglas+02:PNS}
Douglas, N.~G., Arnaboldi, M., Freeman, K.~C., {et~al.} 2002, \pasp, 114, 1234

\bibitem[{{El-Badry} {et~al.}(2018){El-Badry}, Rix, \& Weisz}]{el_badry+18}
{El-Badry}, K., Rix, H.-W., \& Weisz, D.~R. 2018, \apjl, 860, L17

\bibitem[{Gesicki {et~al.}(2018)Gesicki, Zijlstra, \& Miller~Bertolami}]{gesicki+18}
Gesicki, K., Zijlstra, A.~A., \& Miller~Bertolami, M.~M. 2018, \nat, 2, 580

\bibitem[{Harris {et~al.}(2020)Harris, Millman, {van der Walt}, Gommers, Virtanen, Cournapeau, Wieser, Taylor, Berg, Smith, Kern, Picus, Hoyer, {van Kerkwijk}, Brett, Haldane, {del Río}, Wiebe, Peterson, {Gérard-Marchant}, Sheppard, Reddy, Weckesser, Abbasi, Gohlke, \& Oliphant}]{harris+20:numpy}
Harris, C.~R., Millman, K.~J., {van der Walt}, S.~J., {et~al.} 2020, \nat, 585, 357

\bibitem[{Harris \& Zaritsky(2009)}]{harris&zaritsky09}
Harris, J. \& Zaritsky, D. 2009, \aj, 138, 1243

\bibitem[{Hartke {et~al.}(2017)Hartke, Arnaboldi, Longobardi, Gerhard, Freeman, Okamura, \& Nakata}]{hartke+17}
Hartke, J., Arnaboldi, M., Longobardi, A., {et~al.} 2017, \aap, 603, A104

\bibitem[{Henize \& Westerlund(1963)}]{henize&westerlund63}
Henize, K.~G. \& Westerlund, B.~E. 1963, \apj, 137, 747

\bibitem[{Hollands {et~al.}(2024)Hollands, Littlefair, \& Parsons}]{hollands+24}
Hollands, M.~A., Littlefair, S.~P., \& Parsons, S.~G. 2024, \mnras, 527, 9061

\bibitem[{Holy {et~al.}(2022)Holy, Kittisopikul, Wadell, Aschan, Lyon, Lucas, Deits, Mathur, Tambe, Kaisermayer, N5N3, Lihm, Weidner, Brandås, Baldassi, Stocker, {getzze}, MatFi, Baran, Ahn, Chen, Johnson, Gagnon, Millea, Pasquier, Arslan, Karrasch, \& Ranocha}]{holy+:interpolations.jl}
Holy, T., Kittisopikul, M., Wadell, A., {et~al.} 2022, Zenodo

\bibitem[{Hunter(2007)}]{hunter07:matplotlib}
Hunter, J.~D. 2007, Computing in Science and Engineering, 9, 90

\bibitem[{Izotov \& Thuan(2004)}]{izotov&thuan04}
Izotov, Y.~I. \& Thuan, T.~X. 2004, \apj, 602, 200

\bibitem[{Jacoby(1980)}]{jacoby80}
Jacoby, G.~H. 1980, \apjs, 42, 1

\bibitem[{Jacoby(1989)}]{jacoby89:pnlfI}
Jacoby, G.~H. 1989, \apj, 339, 39

\bibitem[{Jacoby \& Ciardullo(2025)}]{jacoby&ciardullo25}
Jacoby, G.~H. \& Ciardullo, R. 2025, \apj, 983, 129

\bibitem[{Jacoby {et~al.}(1990)Jacoby, Ciardullo, \& Ford}]{jacoby+90:pnlfV}
Jacoby, G.~H., Ciardullo, R., \& Ford, H.~C. 1990, \apj, 356, 332

\bibitem[{Jacoby {et~al.}(2024)Jacoby, Ciardullo, Roth, Arnaboldi, \& Weilbacher}]{jacoby+24:II}
Jacoby, G.~H., Ciardullo, R., Roth, M.~M., Arnaboldi, M., \& Weilbacher, P.~M. 2024, \apjs, 271, 40

\bibitem[{Jacoby \& De~Marco(2002)}]{jacoby&de_marco02}
Jacoby, G.~H. \& De~Marco, O. 2002, \aj, 123, 269

\bibitem[{Jimenez {et~al.}(2003)Jimenez, Flynn, MacDonald, \& Gibson}]{jimenez+03}
Jimenez, R., Flynn, C., MacDonald, J., \& Gibson, B.~K. 2003, Science, 299, 1552

\bibitem[{Keszthelyi {et~al.}(2024)Keszthelyi, Puls, Chiaki, Nagakura, {ud-Doula}, Takiwaki, \& Tominaga}]{keszthelyi+24}
Keszthelyi, Z., Puls, J., Chiaki, G., {et~al.} 2024, \mnras, 533, 3457

\bibitem[{Kippenhahn {et~al.}(2013)Kippenhahn, Weigert, \& Weiss}]{kippenhahn+13}
Kippenhahn, R., Weigert, A., \& Weiss, A. 2013, Stellar {{Structure}} and {{Evolution}}, 2nd edn., Astronomy and {{Astrophysics Library}} (Heidelberg: Springer Berlin)

\bibitem[{Kreckel {et~al.}(2017)Kreckel, Groves, Bigiel, Blanc, Kruijssen, Hughes, Schruba, \& Schinnerer}]{kreckel+17}
Kreckel, K., Groves, B., Bigiel, F., {et~al.} 2017, \apj, 834, 174

\bibitem[{Li {et~al.}(2018)Li, Mao, Cappellari, Ge, Long, Li, Mo, Li, Zheng, Bundy, Thomas, Brownstein, Roman~Lopes, Law, \& Drory}]{li+18:gradients}
Li, H., Mao, S., Cappellari, M., {et~al.} 2018, \mnras, 476, 1765

\bibitem[{Longobardi {et~al.}(2013)Longobardi, Arnaboldi, Gerhard, Coccato, Okamura, \& Freeman}]{longobardi+13}
Longobardi, A., Arnaboldi, M., Gerhard, O., {et~al.} 2013, \aap, 558, A42

\bibitem[{Marigo {et~al.}(2020)Marigo, Cummings, Curtis, Kalirai, Chen, Tremblay, {Ramirez-Ruiz}, Bergeron, Bladh, Bressan, Girardi, Pastorelli, Trabucchi, Cheng, Aringer, \& Tio}]{marigo+20}
Marigo, P., Cummings, J.~D., Curtis, J.~L., {et~al.} 2020, \nat, 4, 1102

\bibitem[{Marigo {et~al.}(2004)Marigo, Girardi, Weiss, Groenewegen, \& Chiosi}]{marigo+04:II}
Marigo, P., Girardi, L., Weiss, A., Groenewegen, M. A.~T., \& Chiosi, C. 2004, \aap, 423, 995

\bibitem[{McDonald {et~al.}(2022)McDonald, Davies, \& Beasor}]{mcdonald+22}
McDonald, S. L.~E., Davies, B., \& Beasor, E.~R. 2022, \mnras, 510, 3132

\bibitem[{M{\'e}ndez(2017)}]{mendez17}
M{\'e}ndez, R.~H. 2017, in IAU Symposium, Vol. 323, Planetary Nebulae: Multi-Wavelength Probes of Stellar and Galactic Evolution, eprint: arXiv:1610.08625, 298--302

\bibitem[{M{\'e}ndez {et~al.}(1993)M{\'e}ndez, Kudritzki, Ciardullo, \& Jacoby}]{mendez+93}
M{\'e}ndez, R.~H., Kudritzki, R.~P., Ciardullo, R., \& Jacoby, G.~H. 1993, \aap, 275, 534

\bibitem[{M{\'e}ndez \& Soffner(1997)}]{mendez&soffner97}
M{\'e}ndez, R.~H. \& Soffner, T. 1997, \aap, 321, 898

\bibitem[{M{\'e}ndez {et~al.}(2008)M{\'e}ndez, Teodorescu, Sch{\"o}nberner, Jacob, \& Steffen}]{mendez+08}
M{\'e}ndez, R.~H., Teodorescu, A.~M., Sch{\"o}nberner, D., Jacob, R., \& Steffen, M. 2008, \apj, 681, 325

\bibitem[{Miller~Bertolami(2016)}]{miller_bertolami16}
Miller~Bertolami, M.~M. 2016, \aap, 588, A25

\bibitem[{Nanni {et~al.}(2014)Nanni, Bressan, Marigo, \& Girardi}]{nanni+14}
Nanni, A., Bressan, A., Marigo, P., \& Girardi, L. 2014, \mnras, 438, 2328

\bibitem[{Osterbrock \& Ferland(2006)}]{osterbrock&ferland06}
Osterbrock, D.~E. \& Ferland, G.~J. 2006, Astrophysics of {{Gaseous Nebulae}} and {{Active Galactic Nuclei}} (University Science Books)

\bibitem[{Padovani \& Matteucci(1993)}]{padovani&matteucci93}
Padovani, P. \& Matteucci, F. 1993, \apj, 416, 26

\bibitem[{Pe{\~n}a {et~al.}(2012)Pe{\~n}a, {Reyes-P{\'e}rez}, {Hern{\'a}ndez-Mart{\'i}nez}, \& {P{\'e}rez-Guill{\'e}n}}]{pena+12}
Pe{\~n}a, M., {Reyes-P{\'e}rez}, J., {Hern{\'a}ndez-Mart{\'i}nez}, L., \& {P{\'e}rez-Guill{\'e}n}, M. 2012, \aap, 547, A78

\bibitem[{Pietrinferni {et~al.}(2013)Pietrinferni, Cassisi, Salaris, \& Hidalgo}]{pietrinferni+13}
Pietrinferni, A., Cassisi, S., Salaris, M., \& Hidalgo, S. 2013, \aap, 558, A46

\bibitem[{Pulsoni {et~al.}(2018)Pulsoni, Gerhard, Arnaboldi, Coccato, Longobardi, Napolitano, Moylan, Narayan, Gupta, Burkert, Capaccioli, {Chies-Santos}, Cortesi, Freeman, Kuijken, Merrifield, Romanowsky, \& Tortora}]{pulsoni+18:ePNS}
Pulsoni, C., Gerhard, O., Arnaboldi, M., {et~al.} 2018, \aap, 618, A94

\bibitem[{Reid \& Parker(2010)}]{reid&parker10:III}
Reid, W.~A. \& Parker, Q.~A. 2010, \mnras, 405, 1349

\bibitem[{Rekola {et~al.}(2005)Rekola, Richer, McCall, Valtonen, Kotilainen, \& Flynn}]{rekola+05}
Rekola, R., Richer, M.~G., McCall, M.~L., {et~al.} 2005, \mnras, 361, 330

\bibitem[{Renzini(1981)}]{renzini81}
Renzini, A. 1981, Annales de Physique, 6, 87

\bibitem[{Renzini \& Buzzoni(1986)}]{renzini&buzzoni86}
Renzini, A. \& Buzzoni, A. 1986, in Astrophysics and Space Science Library, Vol. 122, Proceedings of the Fourth Workshop (Erice, Italy: Dordrecht, D. Reidel Publishing Co.), 195--231

\bibitem[{Richer(1993)}]{richer93}
Richer, M.~G. 1993, \apj, 415, 240

\bibitem[{{Rodr{\'i}guez-Gonz{\'a}lez} {et~al.}(2015){Rodr{\'i}guez-Gonz{\'a}lez}, {Hern{\'a}ndez-Mart{\'i}nez}, Esquivel, Raga, Stasi{\'n}ska, Pe{\~n}a, \& Mayya}]{rodriguez_gonzalez+15}
{Rodr{\'i}guez-Gonz{\'a}lez}, A., {Hern{\'a}ndez-Mart{\'i}nez}, L., Esquivel, A., {et~al.} 2015, \aap, 575, A1

\bibitem[{Roth {et~al.}(2021)Roth, Jacoby, Ciardullo, Davis, Chase, \& Weilbacher}]{roth+21:I}
Roth, M.~M., Jacoby, G.~H., Ciardullo, R., {et~al.} 2021, \apj, 916, 21

\bibitem[{Scheuermann {et~al.}(2022)Scheuermann, Kreckel, Anand, Blanc, Congiu, Santoro, Van~Dyk, Barnes, Bigiel, Glover, Groves, Klessen, Kruijssen, Rosolowsky, Schinnerer, Schruba, Watkins, \& Williams}]{scheuermann+22}
Scheuermann, F., Kreckel, K., Anand, G.~S., {et~al.} 2022, \mnras, 511, 6087

\bibitem[{Sch{\"o}nberner(1983)}]{schoenberner83:tracks}
Sch{\"o}nberner, D. 1983, \apj, 272, 708

\bibitem[{Sch{\"o}nberner {et~al.}(2010)Sch{\"o}nberner, Jacob, Sandin, \& Steffen}]{schoenberner+10}
Sch{\"o}nberner, D., Jacob, R., Sandin, C., \& Steffen, M. 2010, \aap, 523, A86

\bibitem[{Sch{\"o}nberner {et~al.}(2007)Sch{\"o}nberner, Jacob, Steffen, \& Sandin}]{schoenberner+07:IV}
Sch{\"o}nberner, D., Jacob, R., Steffen, M., \& Sandin, C. 2007, \aap, 473, 467

\bibitem[{Soemitro {et~al.}(2023)Soemitro, Roth, Weilbacher, Ciardullo, Jacoby, {Monreal-Ibero}, Castro, \& Micheva}]{soemitro+23:IV}
Soemitro, A.~A., Roth, M.~M., Weilbacher, P.~M., {et~al.} 2023, \aap, 671, A142

\bibitem[{Soker(2006)}]{soker06}
Soker, N. 2006, \apj, 640, 966

\bibitem[{Souropanis {et~al.}(2023)Souropanis, Chiotellis, Boumis, Jones, \& Akras}]{souropanis+23}
Souropanis, D., Chiotellis, A., Boumis, P., Jones, D., \& Akras, S. 2023, \mnras, 521, 1808

\bibitem[{Spriggs {et~al.}(2020)Spriggs, Sarzi, Napiwotzki, {Galán-de Anta}, Viaene, Nedelchev, Coccato, Corsini, {de Zeeuw}, {Falcón-Barroso}, Gadotti, Iodice, Lyubenova, {Martín-Navarro}, McDermid, Pinna, {van de Ven}, \& Zhu}]{spriggs+20}
Spriggs, T.~W., Sarzi, M., Napiwotzki, R., {et~al.} 2020, \aap, 637, A62

\bibitem[{Valenzuela {et~al.}(2019)Valenzuela, Méndez, \& Miller~Bertolami}]{valenzuela+19}
Valenzuela, L.~M., Méndez, R.~H., \& Miller~Bertolami, M.~M. 2019, \apj, 887, 65

\bibitem[{Valenzuela {et~al.}(2024)Valenzuela, Remus, Miller~Bertolami, \& Méndez}]{valenzuela+24:iaus}
Valenzuela, L.~M., Remus, R.-S., Miller~Bertolami, M.~M., \& Méndez, R.~H. 2024, in IAU Symposium, Vol. 384, Planetary Nebulae: a Universal Toolbox in the Era of Precision Astrophysics, 1--1

\bibitem[{Vassiliadis \& Wood(1994)}]{vassiliadis&wood94:tracks}
Vassiliadis, E. \& Wood, P.~R. 1994, \apjs, 92, 125

\bibitem[{{von Steiger} \& Zurbuchen(2016)}]{von_steiger&zurbuchen16}
{von Steiger}, R. \& Zurbuchen, T.~H. 2016, \apj, 816, 13

\bibitem[{Yao \& Quataert(2023)}]{yao&quataert23}
Yao, P.~Z. \& Quataert, E. 2023, \apj, 957, 30

\end{thebibliography}

\begin{appendix}

\section{Post-AGB track interpolation}
\label{app:postagb_tracks}

Here we present how the post-AGB tracks provided by \citet{miller_bertolami16} are interpolated (see \cref{sec:postagb_tracks} for the corresponding section in the main text).
For the post-AGB tracks, we take the output tables from \citet{miller_bertolami16}, which each consist of 101~points of equivalent position along the tracks in the Hertzsprung-Russel diagram, with precomputed luminosity, effective temperature, and post-AGB time since a temperature of $\log T_\mathrm{eff} / \si{\kelvin} = 3.85$ was reached. The tables are available for between 12 and 20 final masses for each of the four metallicities $Z = 0.0001$ (12 between $M_\mathrm{final} = 0.525$ and \SI{0.800}{\Msun}), $Z = 0.001$, $Z = 0.01$, and $Z = 0.02$ (each of them have 20 tables for final masses between $M_\mathrm{final} = 0.525$ and \SI{1.000}{\Msun}).
As we follow the approach of \citet{valenzuela+19} in this work to define the post-AGB age as the time since a temperature of $T_\mathrm{eff} = \SI{25000}{\kelvin}$ was reached, the precomputed post-AGB ages are shifted accordingly in the individual tables.
To interpolate the tracks for arbitrary metallicities, final masses, and post-AGB ages, we perform linear interpolations on the logarithmic metallicity and the final mass to find the eight relevant equivalent points on the tracks corresponding to the two adjacent metallicities, final masses, and post-AGB ages. The luminosity and effective temperatures are then determined from these eight points by trilinear interpolation, the result of which can be seen in \cref{fig:postagb_tracks}.

\section{Nebular model recipes}

The nebular model by \citet{valenzuela+19} as described in \cref{sec:pn_model} employs empirical relations for the intensity ratio between $I(5007)$ and $I(\Hbeta)$ and for the absorption factor that accounts for the possibility of optically thin nebulae and their dissipation over time. In this appendix we describe these recipes in more detail to present the final model from \citet{valenzuela+19}, which was built on previous versions by \citet{mendez+93} and \citet{mendez&soffner97}.

\subsection{Intensity ratio recipe}
\label{app:intensity_ratio}

After having obtained $I(\Hbeta)$ assuming a blackbody spectrum based on the luminosity and effective temperature of the central star, the relative intensity of $I(5007)$ is determined empirically. For this we consider three different cases based on $T_\mathrm{eff}$:
\begin{itemize}
    \item $T_\mathrm{eff} \leq \SI{30152}{\kelvin}$: $I(5007)$ is set to zero.
    \item $\SI{30152}{\kelvin} < T_\mathrm{eff} \leq \SI{60000}{\kelvin}$: The intensity is set to $100 I(5007) / I(\Hbeta) = 0.0396 T_\mathrm{eff} / \si{\kelvin} - \num{1194}$ based on the linear relation derived by \citet{mendez+93} in their eq.~5.
    \item $T_\mathrm{eff} > \SI{60000}{\kelvin}$: First, a random value $I_\mathrm{rel}$ is drawn from a normal distribution with a mean of 9.50 and a standard deviation of 3.75, truncated to 0 if a negative value is drawn. As \citet{valenzuela+19} describe, the steep drop-off at the high end of the distribution has to be taken into account to avoid overly high values of $I(5007)$: First, for $I_\mathrm{rel} > 12$ the value is decreased by $((I_\mathrm{rel} - 12) / 8)^{1.5}$. If it is then still larger than 15, it is further decreased by $(I_\mathrm{rel} - 15) / 5$. Finally, if the value is still larger than 18, $I_\mathrm{rel}$ is simply set to a uniformly random value between 18 and 19 and then truncated to a maximum value of 18.5. In a last step, for low central star masses of $M_\mathrm{final} \leq \SI{0.55}{\Msun}$ that are hot and luminous with $\log(L/\Lsun) > 3$ and $T_\mathrm{eff} > \SI{75000}{\kelvin}$, the value of $I_\mathrm{rel}$ is multiplied with 0.4 as described by \citet{valenzuela+19}. This value of $I_\mathrm{rel}$ is finally used to obtain $I(5007)$ by $I(5007) = I_\mathrm{rel} I(\Hbeta)$.
\end{itemize}

\subsection{Absorption factor recipe}
\label{app:absorption_factor}

The absorption factor $\mu$ used by \citet{valenzuela+19} is determined through a linear combination based on central star mass, effective temperature, and post-AGB age relative to reaching the highest combination of effective temperature and luminosity at the knee of the post-AGB tracks. The post-AGB time to the knee is approximated through the analytical relation $t_\mathrm{knee} = 0.0239 M_\mathrm{final}^{-21.82}$. The absorption factor is a value between 0 and 1 that is multiplied with the originally obtained value for $I(5007)$ to account for a lack of optical thickness and nebular dissipation. First, a generator value $\mu_\mathrm{gen}$ is randomly drawn between 0 and 1 from a uniform distribution. For the cooling tracks with post-AGB ages larger than 1.1 times the time to reach the knee, the absorption factor is determined as $\mu = (0.1 + 0.9\mu_\mathrm{gen}^{0.9}) \times (1 - t_\mathrm{post-AGB} / \SI{30000}{\year})$ to account for the nebulae becoming transparent over time. For the heating tracks until the knee, three temperature ranges are considered, where linear interpolation is applied for temperatures between the respective limits:
\begin{itemize}
    \item $\log T_\mathrm{eff} / \si{\kelvin} < 4.51$: For the beginning of the heating track, $\mu$ is set to $\min(0.4 + \mu_\mathrm{gen}, 1)$ to account for generally more opaque PNe.
    \item $4.68 < \log T_\mathrm{eff} / \si{\kelvin} < 4.85$: For low-mass central stars below \SI{0.535}{\Msun}, $\mu = 0.03 + 0.17 \mu_\mathrm{gen}$ is used to obtain overall low values of $\mu$ because of their slow post-AGB evolution, such that the nebulae already dissipate during this time. In contrast, for higher-mass central stars above \SI{0.557}{\Msun}, $\mu = 0.2 + 0.8 (1 - \mu_\mathrm{gen}^2)$ is applied, which allows for high values of $\mu$. Linear interpolation is applied for the mass range in between these mass limits.
    \item $\log T_\mathrm{eff} / \si{\kelvin} > 4.95$: For low-mass central stars below \SI{0.535}{\Msun}, $\mu = 0.03 + 0.17 \mu_\mathrm{gen}$ is again used to account for their slow evolution along the post-AGB tracks. For higher-mass central stars above \SI{0.557}{\Msun}, $\mu = 0.1 + 0.9 \mu_\mathrm{gen}$ is applied, which now leads to lower values of $\mu$ with a higher probability compared to at lower effective temperatures due to the time that has passed. Linear interpolation is again applied for the mass range in between these mass limits.
\end{itemize}
For the time between 1 and 1.1 times the time to reach the knee of the post-AGB track, and for the effective temperatures between the ranges listed above, linear interpolation is applied between the respective obtained values for $\mu$. As a final step, $\mu$ is truncated to the range between \num{e-8} and 1.

\section{Nebular metallicity}
\label{app:nebular_metallicity}

\begin{figure}[h!]
    \centering
    \includegraphics[width=0.98\columnwidth]{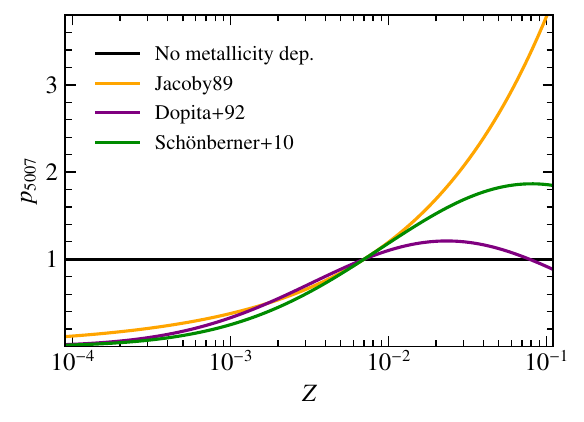}
    \caption{Nebular metallicity correction factor $p_{5007}$ as a function of metallicity for the relations given by \citet{jacoby89:pnlfI}, \citet{dopita+92}, and \citet{schoenberner+10}, as well as for having no metallicity dependence. The relations are set up to reach a value of $p_{5007} = 1$ at the base metallicity $Z_\mathrm{base} = 0.007$.}
    \label{fig:nebular_metallicity}
\end{figure}

In \cref{fig:nebular_metallicity} we show the nebular metallicity correction factor $p_{5007}$, which is multiplied by the obtained \OIII{} fluxes from the model by \citet{valenzuela+19}, as a function of metallicity. The relations are taken from \citet{jacoby89:pnlfI}, \citet{dopita+92}, and \citet{schoenberner+10} and are described in detail in \cref{sec:pn_model}. If no metallicity dependence is taken into account, the flux remains the same with a factor of 1 (black line). As we selected the base metallicity $Z_\mathrm{base} = 0.007$ based on which the corrections are computed, all lines run through $p_{5007}$ at that metallicity.
Because the values of $p_{5007}$ do not lead to as heavily boosted or dimmed \OIII{} fluxes, we selected the relation by \citet{dopita+92} for the fiducial PICS model.

\section{Circumnebular extinction}
\label{app:circumnebular_extinction}

\begin{figure}[h!]
    \centering
    \includegraphics[width=0.98\columnwidth]{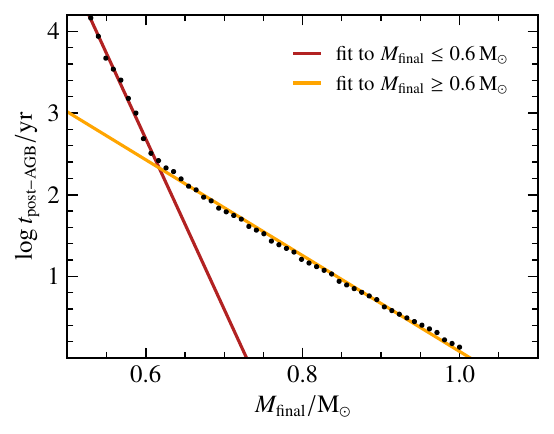}
    \caption{Post-AGB age $t_\mathrm{post-AGB}$ at which PNe at metallicity $Z = 0.01$ reach the brightest possible \OIII{} flux as a function of final mass $M_\mathrm{final}$. The black data points mark the determined values from the model for 50 equally spaced final masses between 0.53 and \SI{1.00}{\Msun}, and the red and orange lines are the linear fits to the values below and above $M_\mathrm{final} = \SI{0.60}{\Msun}$, respectively.}
    \label{fig:brightest_ages}
\end{figure}

The relations for circumnebular extinction by \citet{jacoby&ciardullo25} are only derived from the brightest PNe of \m{31} and the LMC, and they are only dependent on the final stellar mass.
In \cref{sec:circumnebular_extinction}, we describe our choice of using PNe of metallicity $Z = 0.01$ to obtain the post-AGB ages at which the maximal brightness can be reached. Here we present the detailed derivation of this relation.
We first modeled the intrinsic \OIII{} flux evolution throughout the PN lifetime of \SI{30000}{\year} for final masses $M_\mathrm{final} = 0.53$--\SI{1.00}{\Msun} at $Z = 0.01$, taking at each timestep the maximal brightness based on \num{10000} realizations of the nebular model.
For each final mass, we then determined the post-AGB age $t_\mathrm{post-AGB}$ at which the highest flux is reached.
We selected 50 equally spaced values in the final mass range, for which the brightest post-AGB ages are shown in \cref{fig:brightest_ages}, using the stellar evolution relations by \citet{miller_bertolami16} and the nebular model by \citet{valenzuela+19}.

As the relation for the logarithmic brightest post-AGB ages appears to be well-described by a broken-line fit, we fit two linear relations to the final masses smaller and larger than \SI{0.60}{\Msun} (red and orange lines, respectively). Since in our recipe we need to invert this relation to obtain the corresponding final mass for a given post-AGB age, we perform the fits as a function of post-AGB age. The analytical form is given in \cref{eq:brightest_ages}.
This approach ensures that a PN at its maximum brightness experiences the extinction from the derived relation from \citet{jacoby&ciardullo25}, while simultaneously accounting for dust dissipating over time.

\section{Luminosity-specific bright PN numbers}
\label{app:alpha25}

In \cref{sec:normalized_pnlfs} we presented the luminosity-specific PN numbers ($\alpha$) and the visibility lifetimes ($\tau_\mathrm{PN}$) for all PNe with magnitudes brighter than $M(5007) = +3.5$, as obtained for the individual SSPs. This limiting magnitude is based on the assumption that PNe span roughly 8 magnitudes of brightness \citep[e.g.,][]{buzzoni+06}.
However, in practice it is impossible to measure the entire population of PNe in a galaxy, where for most galaxies outside the Local Group one is limited to the brightest couple of magnitudes. For this reason, $\alpha_{2.5}$ has become a relevant measure for the luminosity-specific PN number for those PNe with magnitudes within the brightest two and a half magnitudes of the PNLF \citep[e.g.,][]{jacoby+90:pnlfV, buzzoni+06, longobardi+13, hartke+17}. For the values of each of the SSP PNLFs, we assume a fixed PNLF bright-end cutoff of $M(5007) = -4.5$ and count all PNe brighter with $M(5007) \leq -2.0$.
In \cref{tab:alpha25} we show the values of $\alpha_{2.5}$ for the SSP PNLFs, analogously to \cref{tab:alpha} with the bolometric luminosities obtained through the GALAXEV CB07 SSP models of \citet{bruzual&charlot03:galaxev} using a \citet{chabrier03:imf} IMF.
Compared to $\alpha$, these values are necessarily lower as the magnitude range is smaller.
The values are generally compatible with results from observational work, where typically observed values are on the order of $\alpha_{2.5} \sim \SI{e-8}{\per\Lsun}$ \citep[e.g.,][]{longobardi+13}, which corresponds to values for ages of $t_\mathrm{age} \sim 10$--\SI{12}{\giga\year}, depending on metallicity (see \cref{tab:alpha25}).
The corresponding PN visibility lifetimes $\tau_{\mathrm{PN},2.5}$ are shown in \cref{tab:tau_pn25}, which indicate the time a PN spends being brighter than $M(5007) = -2.0$.
These are also smaller than the values of $\tau_\mathrm{PN}$ (\cref{tab:tau_pn}) by necessity.

\begin{table}[h!]
    \centering
    \caption{Luminosity-specific PN numbers, $\alpha_{2.5}$ in units of \SI{e-7}{\per\Lsun}, of the SSP PNLFs for the brightest PNe with $M(5007) \leq -2.0$. Each row corresponds to the given age $t_\mathrm{age}$ and each column to the given metallicity $Z$.}
    \label{tab:alpha25}
    \begin{tabular}{cS[table-format=1.3,table-auto-round=true]S[table-format=1.3,table-auto-round=true]S[table-format=1.3,table-auto-round=true]S[table-format=1.3,table-auto-round=true]S[table-format=1.3,table-auto-round=true]S[table-format=1.3,table-auto-round=true]}
      \hline
      $t_\mathrm{age}$ & \multicolumn{6}{c}{$Z$} \\
      (\si{\giga\year}) & {0.0001} & {0.001} & {0.01} & {0.02} & {0.04} & {0.08} \\
      \hline\\[-0.2cm]
      0.25 & {--} & {--} & {--} & 9.0e-5 & {--} & {--} \\
      0.5 & {--} & 0.00623 & 0.04943 & 0.03978 & 0.08962 & 0.41479 \\
      1.0 & {--} & 0.10268 & 0.44255 & 0.36418 & 0.25461 & 0.48309 \\
      2.0 & {--} & 0.17273 & 0.80664 & 0.68222 & 0.63114 & 0.38853 \\
      4.0 & {--} & 0.33979 & 1.11571 & 0.81389 & 1.08553 & 0.86904 \\
      6.0 & {--} & 0.58416 & 1.53108 & 1.17289 & 1.0741 & 1.16371 \\
      8.0 & {--} & 0.00097 & 1.08165 & 0.84894 & 1.28046 & 1.43759 \\
      10.0 & {--} & {--} & 0.05151 & 0.21418 & 0.85019 & 0.65512 \\
      12.0 & {--} & {--} & {--} & 0.08694 & 0.13818 & 0.68069 \\
      13.0 & {--} & {--} & {--} & {--} & 0.09405 & 0.57927 \\
      \hline\\[-0.3cm]
      & \multicolumn{6}{c}{$\alpha_{2.5}$ in units of \SI{e-7}{\per\Lsun}} \\
      \hline
    \end{tabular}
\end{table}

\begin{table}[h!]
    \centering
    \caption{PN visibility lifetimes, $\tau_{\mathrm{PN},2.5}$ in units of \SI{e3}{\year}, of the PNe with $M(5007) \leq -2.0$ resulting from the SSPs. Each row corresponds to the given age $t_\mathrm{age}$ and each column to the given metallicity $Z$.}
    \label{tab:tau_pn25}
    \begin{tabular}{cS[table-format=2.2,table-auto-round=true]S[table-format=2.2,table-auto-round=true]S[table-format=2.2,table-auto-round=true]S[table-format=2.2,table-auto-round=true]S[table-format=2.2,table-auto-round=true]S[table-format=2.2,table-auto-round=true]}
      \hline
      $t_\mathrm{age}$ & \multicolumn{6}{c}{$Z$} \\
      (\si{\giga\year}) & {0.0001} & {0.001} & {0.01} & {0.02} & {0.04} & {0.08} \\
      \hline\\[-0.2cm]
      0.25 & {--} & {--} & {--} & 0.0009 & {--} & {--} \\
      0.5 & {--} & 0.05202 & 0.38373 & 0.28356 & 0.7092 & 4.17384 \\
      1.0 & {--} & 0.65832 & 3.09975 & 2.27505 & 1.83417 & 4.10592 \\
      2.0 & {--} & 0.81207 & 3.7014 & 3.42957 & 3.34857 & 2.5998 \\
      4.0 & {--} & 1.96665 & 5.42868 & 4.35117 & 4.44954 & 4.27512 \\
      6.0 & {--} & 3.2628 & 7.85295 & 5.83524 & 5.49366 & 5.14107 \\
      8.0 & {--} & 0.00516 & 5.11185 & 4.11954 & 5.94885 & 5.49399 \\
      10.0 & {--} & {--} & 0.23865 & 1.00464 & 3.92841 & 3.28776 \\
      12.0 & {--} & {--} & {--} & 0.38523 & 0.62004 & 3.27435 \\
      13.0 & {--} & {--} & {--} & {--} & 0.41277 & 2.73108 \\
      \hline\\[-0.3cm]
      & \multicolumn{6}{c}{$\tau_{\mathrm{PN},2.5}$ in units of \SI{e3}{\year}} \\
      \hline
    \end{tabular}
\end{table}

\section{Extended PNLFs}
\label{app:pnlf_extended}

\begin{figure*}
    \centering
    \includegraphics[width=0.98\textwidth]{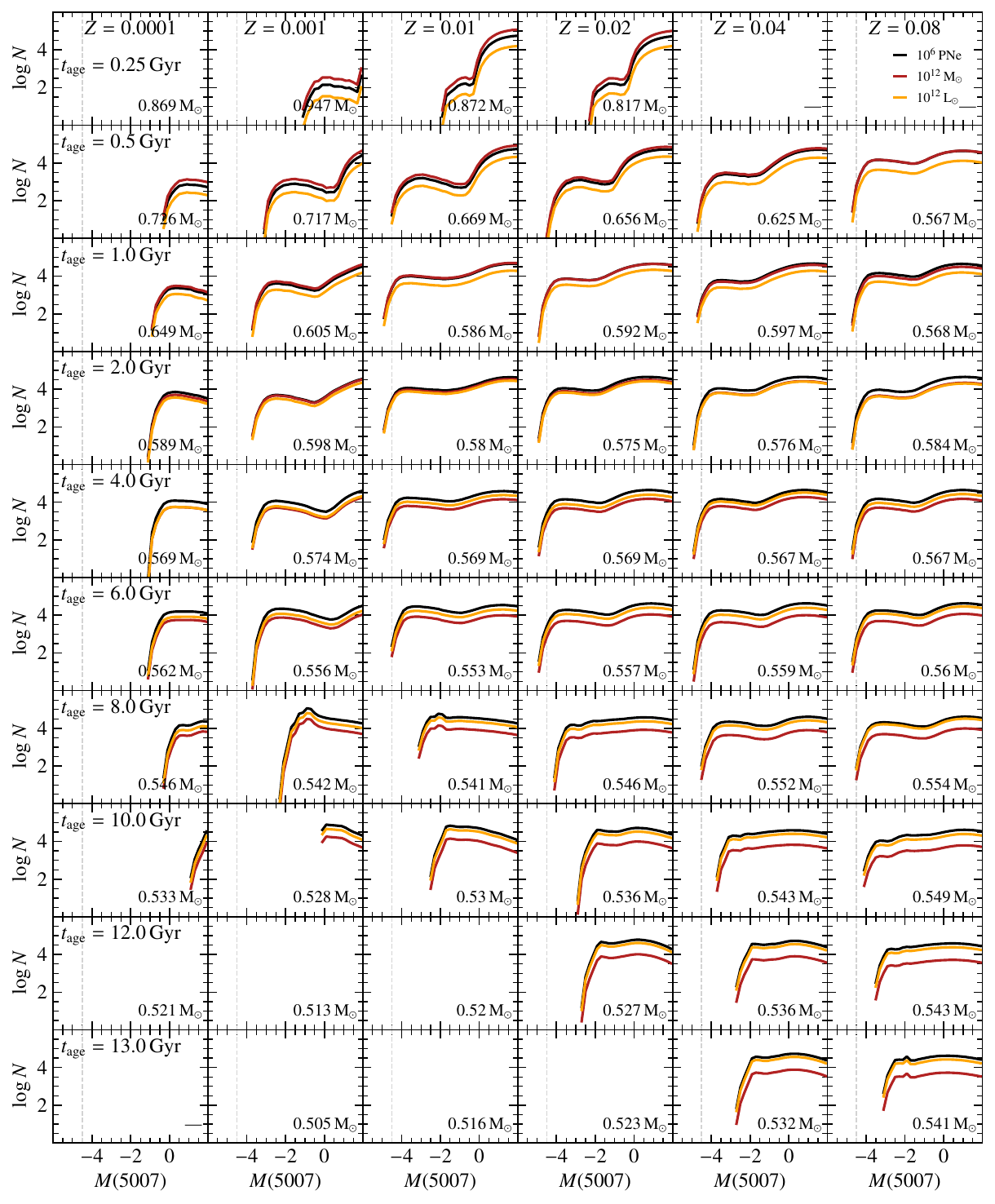}
    \caption{Single stellar population PNLFs depending on the age and metallicity for the fiducial model with the nebular metallicity correction by \citet{dopita+92} and circumnebular extinction by \citet{jacoby&ciardullo25}, normalized to different quantities. This grid of PNLFs shows the same plots as seen in \cref{fig:pnlf_grid}, except that the $\lambda5007$ magnitude range is extended to dimmer PNe by four magnitudes.
    The black lines are the PNLFs obtained from \num{1e6} stars of the respective final mass with equally spaced post-AGB ages between 0~and \SI{30000}{\year}.
    The red PNLFs are normalized to a total mass of \SI{e12}{\Msun} assuming a Chabrier IMF \citep{chabrier03:imf}.
    The orange PNLFs are normalized to a total bolometric luminosity of \SI{e12}{\Lsun} based on the GALAXEV CB07 SSP models by \citet{bruzual&charlot03:galaxev} using a Chabrier IMF.
    }
    \label{fig:pnlf_grid_extended}
\end{figure*}

In \cref{fig:pnlf_grid_extended} we show the same set of SSP PNLFs as in \cref{fig:pnlf_grid} for a grid of different ages and metallicities, just increasing the plotted magnitude range to fainter PNe at $M(5007) = +2$. Here it is much more obvious that most of the individual PNLFs drop towards fainter PNe and many in the young to intermediate age range ($t_\mathrm{age} \lesssim \SI{6}{\giga\year}$) then again have a rising number of PNe towards the faintest magnitudes. This \enquote{camel} shape has previously been found in both observations and simulations of PNe, in particular in younger systems \citep[e.g.,][]{jacoby&de_marco02, reid&parker10:III, valenzuela+19}, where the dip in the SMC occurs at a slightly dimmer magnitude of $M(5007) \approx -0.5$ \citep{jacoby&de_marco02}. This is consistent with the grid of PNLFs shown here, where the SSP models in \cref{fig:pnlf_grid_extended} show that this shape can also arise within a pure SSP and it is not dependent on the correct superposition of different SSP PNLFs. In contrast, it is possible to wash out the camel shape by combining multiple SSP PNLFs as their peaks are located at different magnitudes and their shapes also differ in declining and increasing slope.
The extreme rise in PN number at dim magnitudes for the youngest stellar populations of $t_\mathrm{age} \lesssim \SI{0.5}{\giga\year}$ arises from the circumnebular extinction decreasing with post-AGB age. This means that the extinction affects the intrinsically brightest PNe in the early post-AGB phase the strongest. In contrast, the older, but much dimmer PNe are hardly affected by the extinction. The PNLF is thereby effectively squashed towards dim magnitudes.

\end{appendix}

\end{document}